\documentclass[aps,prb,superscriptaddress,11pt]{revtex4-2}

\usepackage{amsmath,amssymb}
\usepackage{graphicx}
\usepackage[compact]{titlesec} 
\usepackage[headheight=0pt,top=1.5cm,margin=2cm]{geometry}

\usepackage{pifont}    

\usepackage{wrapfig}
\usepackage{float}
\makeatletter
\@ifundefined{newfloat@ltx}{}{%
  \let\newfloat\newfloat@ltx
}
\makeatother
\usepackage{algorithm}
\usepackage{algorithmic}

\usepackage[toc,page]{appendix}
\usepackage{adjustbox} 

\usepackage{enumitem}
\usepackage{caption}
\usepackage{booktabs}  
\usepackage{array}     
\usepackage{subcaption}

\usepackage{overpic}

\usepackage[braket, qm]{qcircuit}
\usepackage{listings}

\usepackage{xcolor}

\usepackage{setspace}
\usepackage{tikz}
\usepackage{braket}

\usepackage{titlesec}
\titleformat{\chapter}[display]
  {\normalfont\bfseries\Huge}
  {}
  {0pt}
  {}

\usepackage{siunitx}

\begin{document}

\pagenumbering{roman}




\title{Quantum Encoding of Structured Data with Matrix Product States}

\author{Josh Green}
 \email{josh.green@uwa.edu.au}
 \affiliation{Centre for Quantum Information, 	
Simulation and Algorithms, School of Physics, Mathematics and Computing, The University of Western Australia, Perth, Australia}
\author{Jingbo Wang}%
 \email{jingbo.wang@uwa.edu.au}
 \affiliation{Centre for Quantum Information, 	
Simulation and Algorithms, School of Physics, Mathematics and Computing, The University of Western Australia, Perth, Australia}



\maketitle
    
    \begin{center}
    \textbf{Abstract}
    \end{center}
    
    The amplitude encoding of an arbitrary $n$-qubit state vector requires $\Omega(2^n)$ gate operations, owing to the exponential dimension of the Hilbert space. We can, however, form dimensionality-reduced representations of quantum states using matrix product states (MPS). In this article, we illustrate that MPS techniques enable the preparation of quantum states representative of functions with complexity up to low-degree piecewise polynomials via shallow-depth quantum circuits with accuracy exceeding 99.99\%. We extend these results to the approximate amplitude encoding of pixel values. We showcase this approach by efficiently preparing a $128\times 128$ ChestMNIST medical image (https://medmnist.com/) on 14 qubits with fidelity exceeding 99.2\% on a circuit with a total depth of just 425 single-qubit rotation and CNOT gates. 

\pagenumbering{arabic}

\vspace{0.5cm}

\section*{1. Introduction}

A crucial component of many quantum applications is state preparation, which involves designing quantum circuits to encode classical information into quantum amplitudes. Quantum state preparation is an imperative subroutine for machine learning \cite{peralgarcia,Cerezo_challenges_and_opportunities_in_QML,Guala2023_qml_image,Landman_2022_qml_image,compressed_prep_for_QML}, the HHL algorithm \cite{harrow_2009_hhl,childs2,liu}, quantum chemistry simulation \cite{dong_ground_state_prep_qet,Low2017_chuang_signal_processing_hamiltonian,childs,motta,shang2023_chemistry_simulation}, computational finance \cite{woerner2019finance},  Monte Carlo simulation \cite{montonaro,Vazquez}, among others. These applications represent some of the most extensively studied and impactful areas in quantum computation.

The core incongruity of quantum state preparation is that the unitary transformation to prepare an arbitrary $n$-qubit state has $\mathcal{O}(2^n)$ real degrees of freedom, which cannot be accurately decomposed into $\mathcal{O}(\text{poly}(n))$ one- and two-qubit quantum gates in general \cite{mottonen2005, mottonen2004,ChenWang2013,LokeWang2016}. Therefore, the design of efficient quantum circuits for the state preparation of any arbitrary target state requires the assumption that the target state has some form of ``polynomial-sized'' intrinsic structure that can be effectively exploited. Moreover, many of the foundational techniques for state preparation -- such as the Grover-Rudolph algorithm -- rely on resource-intensive circuits for quantum arithmetic \cite{gr,mcardle2022quantumstatepreparationcoherent}, or on variational quantum circuits that are prone to barren plateaus during training \cite{larocca2024reviewbarrenplateausvariational,cerezo_cost_function_dependent_BP}.

The matrix product state (MPS) is a one-dimensional tensor network that can approximate quantum states using a tunable number of parameters. Compressed MPS representations accurately represent quantum states characterised by exponentially decaying Schmidt coefficients \cite{orus2014}. The MPS format can be effectively viewed as a dimensionality reduction technique that exploits the limited entanglement structure in specific quantum states. There are many well-established methods of mapping MPS into shallow-depth quantum circuits \cite{schon_sequential_MPS_generation,ran,rudolph2022decompositionmatrixproductstates,malz_mps_encoding,constant_depth_mps_smith,iqbal_2022_mps,BenDov2024}. However, the error introduced in approximating a quantum state with an MPS lacks rigorous error bounds in general. For this reason, it remains theoretically unclear to what extent MPS-based state preparation can be utilised for the state preparation of practical quantum applications. Additionally, while all bond-$\chi$ MPS can be expressed by $\mathcal{O}(n\chi^2)$-depth circuits \cite{malz_mps_encoding}, the specific circuit depth required to achieve a high-quality approximation of a target state is problem-dependent.

We adopt the classically efficient Matrix Product Disentangler (MPD) algorithm introduced by Shi-Ju Ran in 2020 \cite{ran}. The MPD algorithm computes a quantum circuit of $L$ sequential layers of 2-qubit unitaries to approximate the MPS. While this algorithm is efficient, the fidelity of the MPS approximation for a given number of circuit layers $L$ can be increased through tensor network optimisation (TNO). The circuit parameters computed by the MPD algorithm can be used as the initial parameters for TNO, which significantly outperforms comparative TNO schemes that lack a high-quality parameter initialisation. A similar scheme was explored in \cite{rudolph2022decompositionmatrixproductstates}, which considered various approaches to the optimisation component. 
To balance computational costs with performance, we choose to adopt a global optimisation algorithm initialised by the MPD algorithm
\cite{rudolph2022decompositionmatrixproductstates}. For clarity, we denote this algorithm as MPD+TNO in this paper. 

Motivated by theoretical bounds on the entanglement entropy in quantum states representative of discretised functions \cite{holmes_smooth_diff_functions,iaconis}, previous research has identified the MPD algorithm as a viable approach for the state preparation of smooth functions. We first validate these results through an empirical study into low-rank MPS approximations of a broad class of elementary and polynomial functions, including those with irregular and discontinuous characteristics. We then demonstrate that the addition of TNO to the MPD scheme significantly broadens the class of functions that can be efficiently prepared via shallow-depth circuits. This broadened class of functions includes piecewise Chebyshev polynomials, the root and logarithmic functions, and piecewise polynomials with limited discontinuities. The efficient preparation of these functions has immediate applications, such as in quantum finance \cite{Rebentrost_2018_QMC_optionpricing,yusen_portfolio_optimisation}. Quantum states representative of these simple functions obey an area law of entanglement, which for one-dimensional systems implies that the entanglement entropy is independent of system size. Subsequently, our results scale to large numbers of qubits. 

Next, we explore the application of the MPD and MPD+TNO algorithms for approximate quantum image encoding. Previous research has identified that sequential MPS circuits were surprisingly effective in preparing states representative of the Fourier coefficients of Fashion-MNIST images \cite{Jobst2024efficientmps}. On the other hand, a recent empirical study detailed that MPS are unlikely to be sufficiently expressive to capture the entanglement entropy scaling of more realistic natural image datasets \cite{lu_tn_efficient_descriptions_classical_data}. We adopt the $128\times 128$ ChestMNIST dataset \cite{medmnistv1,medmnistv2} as a case study and illustrate that the MPS-based preparation techniques can encode these images with fidelity exceeding 99.0\%. Adding TNO to the MPD algorithm significantly reduces the circuit depth of the computed quantum circuits. For instance, the MPD algorithm obtains a 99.4\% fidelity on the image encoding problem with 100 layers, but a similar fidelity can be obtained with just 50 layers when TNO is used. Enabling the approximate encoding of structured data such as images and discretised functions in linear-depth circuits of 1- and 2-qubit gates, without any ancillas, we identify MPS-based encoding as a leading candidate for simple near-term state preparation tasks. We also clearly establish that the accurate encoding of more complex data structures necessitates improved state preparation techniques, such as those leveraging more powerful tensor network structures. 

\section*{2. Background and Motivation}

This section outlines the conventional methods and confounding challenges of designing general and efficient quantum circuits for state preparation. This serves as a condensed review and critique of the foundational literature in quantum state preparation, centred around comparisons with traditional methods for preparing structured states, which motivates the adoption of MPS-based encoding techniques in the proceeding sections.

\subsection*{2.1 Amplitude Encoding and Quantum State Preparation}
Let $\mathbf{a} = (a_0, \dots, a_{N-1})^T \in \mathbb{C}^N$ be a vector of $N = 2^n$ complex numbers, where $n$ denotes the number of qubits in the system. State preparation aims to encode this vector into the amplitudes of a quantum state wavefunction. The state $\ket{\psi}$, known as the target state, is given by:
\begin{equation}
\ket{\psi} = \frac{1}{\|\mathbf{a}\|_2} \sum_{i=0}^{N-1} a_i \ket{i}
\end{equation}
where $\|\mathbf{a}\|_2$, the Euclidean norm of $\mathbf{a}$, is the normalisation coefficient. $\ket{\psi}$ is prepared by:
\begin{equation}
U_S\ket{0}^{\otimes n} = \ket{\psi'}
\end{equation}
where $U_S\in U(N)$ is the so-called state preparation operator, $\ket{0}^{\otimes n}$ is the initial state, and $\ket{\psi'}$ is an approximation of the target state. We aim to maximise the fidelity $F = |\langle \psi | \psi' \rangle|^2$ with at most $\mathcal{O}(\text{poly}(n))$ circuit depth.

The amplitude encoding of $2^n$ unstructured points requires a one- and two-qubit gate complexity of $\Omega(2^n)$ \cite{mottonen2005}. This stems from the fact that the most general unitary operation mapping the zero state $\ket{0}^{\otimes n}$ to another quantum state has $2^{n+1}-2$ real degrees of freedom, excluding the degrees of freedom associated with the global phase and normalisation constant~\cite{mottonen2004,ChenWang2013,LokeWang2016}. Considering that there is no currently feasible means of maintaining coherence long enough for practical QRAM~\cite{Giovannetti2008}, the issue of inefficient quantum state preparation is exacerbated by the requirement to encode the classical information for every measurement of the output register. For example, it has been argued that the exponential speed-up for qPCA~\cite{Lloyd2014} is derived from state preparation assumptions of QRAM, and there is no viable means to efficiently encode a dense vector $\boldsymbol{b}$ to solve the linear system $A \boldsymbol{x} = \boldsymbol{b}$ without predominating the overall complexity of the HHL algorithm~\cite{harrow_2009_hhl}. 

This highlights critical limitations for a wide range of quantum computing applications, including quantum machine learning, quantum image processing, and quantum signal processing. These fields often rely on efficient quantum state preparation to encode significant classical data into quantum systems. Without scalable and practical solutions to these challenges, the computational overhead required for data encoding could negate the potential advantages of quantum algorithms. Addressing these limitations is therefore essential to unlocking the full potential of quantum technologies in solving real-world problems.

In recent years, a workaround to the exponential circuit depth of arbitrary quantum state preparation has gained traction, namely the space-time trade-off approach \cite{gui_spacetime,zhang_low_depth,zhang_optimal,Araujo_2021_divideandconquer,iten_isometries_encoding}. This approach exploits quantum circuits that effectively \textit{parallelises} $\mathcal{O}(2^n)$ operations using $\mathcal{O}(2^n)$ ancilla qubits, achieving an optimal linear circuit depth of $\mathcal{O}(n)$ \cite{gui_spacetime,zhang_low_depth}. However, this method necessitates exponentially large quantum computers capable of arbitrarily long-range, non-local gate operations, which is unrealistic. Decomposing this circuit into local operations retrieves an exponential asymptotic scaling of circuit depth. 


We are ultimately compelled to assume some form of innate \textit{structure} in the target state that can be exploited for efficient state preparation. The first form of structure to consider is the assumption that the target vector is $m$-sparse, containing at most $m$ non-zero elements, where $m\ll 2^n$. Numerous efficient quantum algorithms have been developed specifically to handle the preparation of sparse states \cite{ramacciotti_2023_sparse_GR,Malvetti_2021_sparse_isom,gleinig_2021_sparse,Feniou_2024_sparse_variationalchem,deVeras_2022_sparse_qram}. However, this approach offers a trivial resolution to the dimensionality burden through the restrictive assumption of a sub-exponential number of non-zero elements in the target state vector. This motivates a consequential question: What kind of non-trivial structures can be leveraged? Furthermore, how complex can these structures be while allowixeng for efficient state preparation? As delineated in Figure 1, we aspire to characterise aspects of the transition period that may enable efficient quantum state preparation. As the target state becomes increasingly complex, it will start to resemble unstructured data, which will then forbid an efficient and accurate $\mathcal{O}(\text{poly}(n))$ quantum circuit to encode it.
\vspace{-0.3cm}
\begin{figure}[!h]
    \centering
    \includegraphics[trim=0cm 0cm 0cm 0cm, clip, width=\textwidth]{Misc_Plots/Entropy_Transition.png}
    \caption{The transition from a highly structured smooth function target vector with low entanglement entropy to fitting a smooth function to an arbitrary unstructured state vector.}
    \label{fig:function_transition}
\end{figure}

\vspace{-0.8cm}
\subsection*{2.2 Existing Approaches to  State Preparation}
\subsubsection*{The Grover-Rudolph Algorithm}
\begin{figure}[H]
    \centering
    \begin{subfigure}[b]{0.60\textwidth}
        \centering
        \includegraphics[trim=0cm 0cm 0cm 0cm, clip, width=\textwidth]{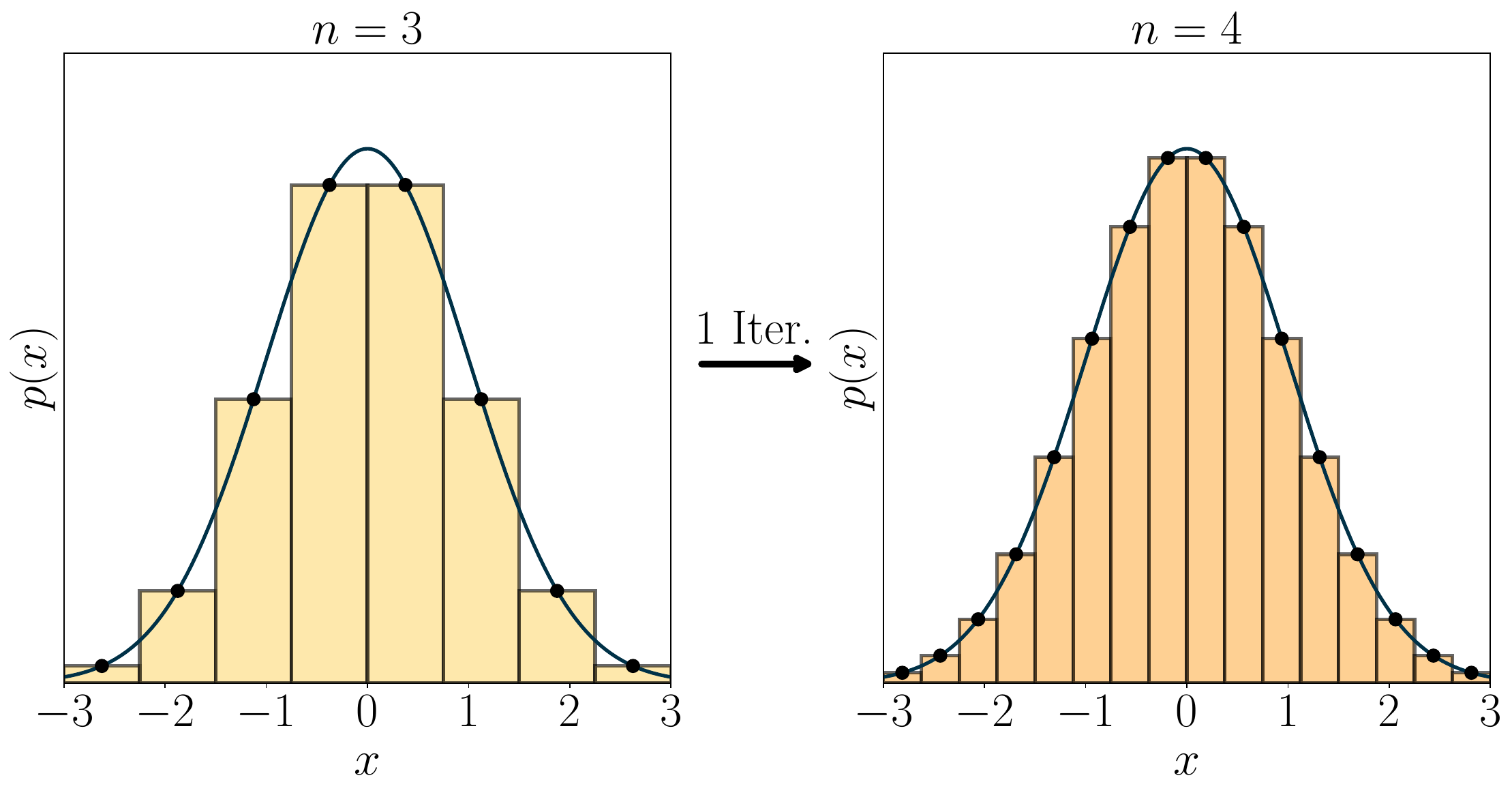}
        \centering
        \caption{}
        \label{fig:grover_doubling}
    \end{subfigure}%
    \hfill
   \begin{subfigure}[b]{0.37\textwidth}
        \centering
        \includegraphics[trim=0.4cm -1.5cm 0.2cm 0cm, clip, width=\textwidth]{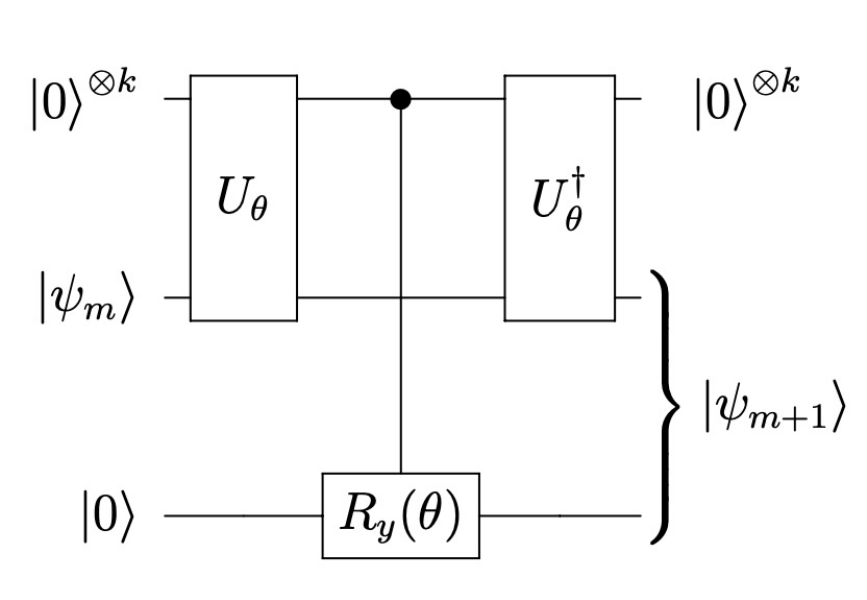}
        \centering
        \caption{}
        \label{fig:grover_doubling}
    \end{subfigure}%
    \caption{The Grover-Rudolph Algorithm: (a) The resolution of the target state doubles with each Grover-Rudolph iteration and (b) The circuit for the $m^{\text{th}}$ iteration of the algorithm.}
    \label{fig:combined_grover}
\end{figure}

A foundational quantum state preparation algorithm is the Grover-Rudolph algorithm, first proposed by Zalka in 1998 \cite{zalka} and independently rediscovered by its namesake in 2002 \cite{gr}. The Grover-Rudolph algorithm is a coarse-graining scheme designed to prepare log-concave probability distributions. We begin with a discretisation of the target state across $m<n$ qubits,
\begin{equation}
    \ket{\psi_m} = \sum_{i=0}^{2^m-1} \sqrt{p_i^{(m)}} \ket{i}_m
\end{equation}
where $p(x)$ is log-concave probability distribution uniformly discretised across $2^m$ point. Next we compute a set of rotation angles $\boldsymbol{\theta} = \{\theta_i\}_{i=0}^{2^m}$, in quantum parallel, using $k$ ancilla qubits,
\begin{equation}
    U_\theta: \sqrt{p_i^{(m)}} \ket{i}_m \ket{0...0}_k \longrightarrow \sqrt{p_i^{(m)}} \ket{i}_m \ket{\theta_i}_k
\end{equation}
where 
$U_\theta$ is a unitary oracle,  
$\theta_i := 2\arccos{\sqrt{f(i)}}$ is a $k$-bit approximation of the $i^{th}$ rotation angle, and 
\begin{equation}
f(i) = \frac{\int_{x_L^i}^{(x_R^i-x_L^i)/2} p(x) \, dx}{\int_{x_L}^{x_R} p(x) \, dx}
    \label{eq:f_in_parallel}
\end{equation}
is the probability that, given $x$ lies in the $i^{th}$ region, it also lies in the left half of this region. 
Here $x_L^i$ and $x_R^i$ are the left and right boundaries of region $i$. Then, applying a controlled-$R_y$ rotation to an ancilla qubit yields the following state:
\begin{equation}
\ket{\psi_{m+1}'} = \sum_{i=0}^{2^{m}-1} \sqrt{p_i^{(m)}} \ket{i}_m \ket{\theta_i}_k (\cos{\theta_i}\ket{0}+\sin{\theta_i}\ket{1})_1 ~.
\end{equation}
Uncomputing with $U_{\theta}^{\dagger}$ disentangles the first register into $\ket{0}^{\otimes k}$, which can be reused for subsequent computations. Now absorbing the state of the ancilla qubit into the second register, we are left with the state:
\begin{equation}
\ket{\psi_{m+1}} = \sum_{i=0}^{2^{m+1}-1} \sqrt{p_i^{(m+1)}} \ket{i}
\end{equation}
which is an $(m+1)$-qubit discretisation of the function with half the discretisation error, as depicted in Figure~\ref{fig:combined_grover}. Fundamentally, we can begin this process with a trivial 1-qubit discretisation of the function. Repeating the process $n-1$ times will yield the target state $\ket{\psi_n}$ to an exponentially small error of $\mathcal{O}(2^{-n})$.

\vspace{-0.2cm}

If the target probability distribution has well-defined integrals, then numerical integration can be avoided in the computation of the rotation angles in Equation~\ref{eq:f_in_parallel}. This enables a more direct arithmetic approach to computing the rotation angles in parallel. Designing circuits for quantum arithmetic is non-trivial, partly due to the requirement that the computation be reversible. This can result in deep circuits with 100's of ancilla qubits \cite{haner_arithmetic}. Moreover, the design of circuits for each specific function is complicated, often requiring some form of optimisation to minimise resource costs. This alludes to a critical underlying issue with the widespread assumption that arbitrary functions can be applied to the basis states of quantum computers: While theoretically efficient, the practical implementation of function-based transformations in the computational basis is limited by the lack of pre-existing circuitry for complex operations. 

Generally, the target probability distribution will not have a well-defined analytic integral. This implies that the integrals in Equation~\ref{eq:f_in_parallel} must be computed using some form of numerical integration, such as classical Monte Carlo simulation in quantum parallel. The error in such an approach scales as $\mathcal{O}(\frac{1}{\sqrt{N}})$ where $N$ is the number of samples, an additional error, which eliminates the quadratic speedup of quantum Monte Carlo integration when the Grover-Rudolph algorithm is used for state preparation \cite{herbert_no_speedup_with_GR_QMCI}. This is a severe limitation for applications in quantum finance, many of which depend on the quadratic speedup of quantum Monte Carlo integration \cite{Rebentrost_2018_QMC_optionpricing}. In fact, this quadratic speedup is expected to be lost in \textit{any} state preparation regime utilising classical Monte Carlo simulation in its computation. 
The circuitry to implement such a routine is not readily available and could be expected to accumulate significant resource costs. Hence, the Grover-Rudolph algorithm is only feasibly achievable when the rotation angles can be easily computed in parallel and does not currently appear to be an effective option for near-term quantum state preparation.

\subsubsection*{Digital-to-Analog Encoding}
Digital-to-analog encoding refers to loading relevant state vector information into the computational basis of the quantum system (digital) before applying a unitary transformation to transfer this information into the amplitudes of the system (analog) \cite{ grover_blackbox2000,mitarai_analog_digital_nonlinear, sanders_no_arithmetic_blackbox,bausch_fast_blackbox,
laneve2023robustblackboxquantumstatepreparation}. This is the oracle unitary:
\begin{equation}
    U_{D\rightarrow A}: \ket{j}_n \ket{a_j}_k \rightarrow a_j \ket{j}_n
\end{equation}
where $a_j$ is a $k$-bit approximation of the $j^{\text{th}}$ element of the normalised target vector $\mathbf{a}$. The foundational algorithm to implement $U_{D\rightarrow A}$ was introduced by Grover in 2000 \cite{grover_blackbox2000} with an asymptotic complexity of $\mathcal{O}(\sqrt{N})$, which is a quadratic speed up over arbitrary state preparation scaling as $\mathcal{O}(N)$. Much like the Grover-Rudolph algorithm, Grover's approach requires the computation of $\arcsin(x)$ via quantum arithmetic in computing the rotation angles of the $U_{D\rightarrow A}$ oracle. Using a non-arithmetic comparison operator in its place was proven to reduce the gate complexity by $\sim 10^2$ with the addition of $\mathcal{O}(n)$ ancilla qubits \cite{sanders_no_arithmetic_blackbox}. The gate complexity and ancilla requirement can be further decreased if we assume knowledge of the average bit-weight of the target vector \cite{bausch_fast_blackbox}.

These encoding schemes are commonly referred to as `black-box' schemes, owing to the assumed existence of an amplitude-loading oracle:
\begin{equation}
    U_A: \ket{j}_n\ket{0}_k \rightarrow \ket{j}_n\ket{a_j}_k
\end{equation} 
or some closely related oracle that prepares the $k$-bit-approximated vector elements in the computational basis. The lack of specification of the gate implementation of this oracle could be useful for quantum machine learning tasks where the inputs are derived directly from a quantum process and are not known classically \cite{Huang_2022_advantageinlearningfromexperiments}. Alternatively, $U_A$ can be implemented efficiently if $\mathbf{a}$ is sparse \cite{Malvetti_2021_sparse_isom}. In general, however, an efficient, practical implementation of $U_A$ requires the assumption that the target vector is a discretised function, in which case the amplitude-loading oracle can be reformulated as:
\begin{equation}
    U_A: \ket{x_j}_n\ket{0}_k \rightarrow \ket{x_j}_n\ket{f(x_j)}_k
    \label{eq:function_load_oracles}
\end{equation}
where $f(x_j)$ represents a $k$-bit approximation of $f: \mathbb{R} \rightarrow \mathbb{R}$ evaluated at the point $x_j$. 

\subsubsection*{Variational Algorithms for State Preparation}
Machine learning-based variational quantum algorithms (VQAs) have been widely applied for state preparation of probability distributions \cite{zoufal_qgan} and quantum chemistry \cite{Kuzmin2020variationalquantum_data_buses_state_prep,gard2020efficient_symmetrypreserving_state_prep_for_variational_QE,theory_of_variational_quantum_groundstate_preparation,variational_quantum_gibbs_state_prep,variational_gibbs_state_prep_NISQ}. VQAs train a parameterised quantum circuit (PQC) to extremise a problem-specific cost function. While VQAs generate (NISQ-friendly) shallow circuits with minimal ancilla qubits \cite{Bharti_2022_NISQ_algorithms,Cerezo_2021_VQAs}, they suffer from critical trainability problems derived from the near-ubiquitous barren plateau phenomena and local minima in the optimisation landscape \cite{Anschuetz2022_VQAs_swamped_traps,cerezo_cost_function_dependent_BP}. The barren plateau is formally defined by the condition:
\begin{equation}
    \text{Var}\left(C(\boldsymbol{\theta})\right) \leq b^n \quad \text{for} \quad 0 < b < 1 ~.
\end{equation}
This implies the exponential suppression of the variance in the cost function as the number of qubits $n$ in the system grows. In turn, cost function gradients decay exponentially, resulting in severe inefficiency in minimising $C(\theta)$ for large $n$ \cite{larocca2024reviewbarrenplateausvariational, cerezo_cost_function_dependent_BP, Cerezo_challenges_and_opportunities_in_QML}. This is not to say that the variance in the cost function is zero everywhere: the barren plateaus typically emerge because of a severe concentration of cost function variance in an exponentially small region of the parameter landscape, which is called the narrow gorge \cite{larocca2024reviewbarrenplateausvariational}. The remaining parameter landscape is featureless and flat. 

Barren plateaus can be avoided in special cases where a \textit{local} cost function is utilised, and the circuit depth of the PQC is sufficiently shallow, i.e. less than $\mathcal{O}(\text{poly(log}(n)))$ \cite{cerezo_cost_function_dependent_BP}. The shallow depth of these circuits constrains their expressibility, i.e. they can only express states belonging to a small subset of the state space. It has recently been suggested that VQAs with such limited expressibility are likely \textit{classically efficient} to train in any case, eliminating the potential quantum advantage of VQAs \cite{cerezo2024doesprovableabsencebarren}. This is not a severe restriction for applying VQAs to state preparation since the quantum advantage is anticipated to emerge in the subsequent quantum algorithm rather than from state preparation itself. Nonetheless, VQA-based state preparation is heavily constrained in terms of what it can achieve. 

\begin{table}[!htbp]
\centering
\small
\setlength{\tabcolsep}{4pt}
\renewcommand{\arraystretch}{1.15}

\begin{tabular}{|c|c|c|c|c|}
\hline
\parbox[t]{3cm}{\raggedright\textbf{Method}} &
\textbf{Depth} &
\textbf{Ancillas} &
\textbf{Avoids Arithmetic} &
\parbox[t]{5cm}{\raggedright\textbf{Scope}} \\
\hline

\parbox[t]{3cm}{\raggedright Space Efficient \cite{mottonen2005}} &
$\mathcal{O}(2^n)$ & 0 & \ding{51} &
\parbox[t]{5cm}{\raggedright Any $n$-qubit state. Exponential depth.} \\
\hline

\parbox[t]{3cm}{\raggedright Time Efficient \cite{gui_spacetime}} &
$\mathcal{O}(n)$ & $\mathcal{O}(2^n)$ & \ding{51} &
\parbox[t]{5cm}{\raggedright Any $n$-qubit state. Exponential ancillas.} \\
\hline

\parbox[t]{3cm}{\raggedright Grover-Rudolph \cite{gr}} &
$\mathcal{O}(\mathrm{poly}(n))$ & \textit{Varies}\textsuperscript{*} & \ding{55} &
\parbox[t]{5cm}{\raggedright Functions with $\mathcal{O}(\mathrm{poly}(n))$ arithmetic circuits.} \\
\hline

\parbox[t]{3cm}{\raggedright Grover Black-Box \cite{grover_blackbox2000}} &
$\mathcal{O}(\sqrt{2^n})$\textsuperscript{\ensuremath{\dagger}} &
\textit{Varies}\textsuperscript{*} & \ding{55} &
\parbox[t]{5cm}{\raggedright Functions with $\mathcal{O}(\mathrm{poly}(n))$ arithmetic circuits.} \\
\hline

\parbox[t]{3cm}{\raggedright Variational Quantum Algorithms} &
\textit{Varies} & \textit{Limited} & \ding{51} &
\parbox[t]{5cm}{\raggedright Prone to barren plateaus and local minima.} \\
\hline
\end{tabular}

\caption{\label{tab:quantum_algorithms_comparison}
A summary of (non-MPS) quantum state preparation methodologies.
\textsuperscript{*}The number of ancilla qubits is generally dominated by quantum arithmetic requirements, which can be substantial~\cite{haner_arithmetic}.
\textsuperscript{\ensuremath{\dagger}}Worst-case rounds of amplitude amplification.}
\end{table}

\section*{3. Matrix Product State Encoding}

In the previous section, we outlined that state preparation techniques dependent on deep VQAs without efficient pre-training generally suffer from barren plateaus. We also noted that the assumption of shallow-depth circuits for quantum arithmetic may require many ancilla qubits and is likely not a permissible approach in the near term. In contrast, MPS-based state preparation techniques have recently become popularised as a resource-efficient approach that avoids the pitfalls of VQAs and quantum arithmetic \cite{ran,rudolph2022decompositionmatrixproductstates,malz_mps_encoding,BenDov2024,constant_depth_mps_smith,iqbal_2022_mps}. This section introduces the MPS format alongside the MPD and MPD+TNO algorithms.

\subsection*{3.1 Formulation}

\begin{figure}[H]
    \centering
    \includegraphics[trim=0cm 3.4cm 0cm 4cm, clip, width=\textwidth]{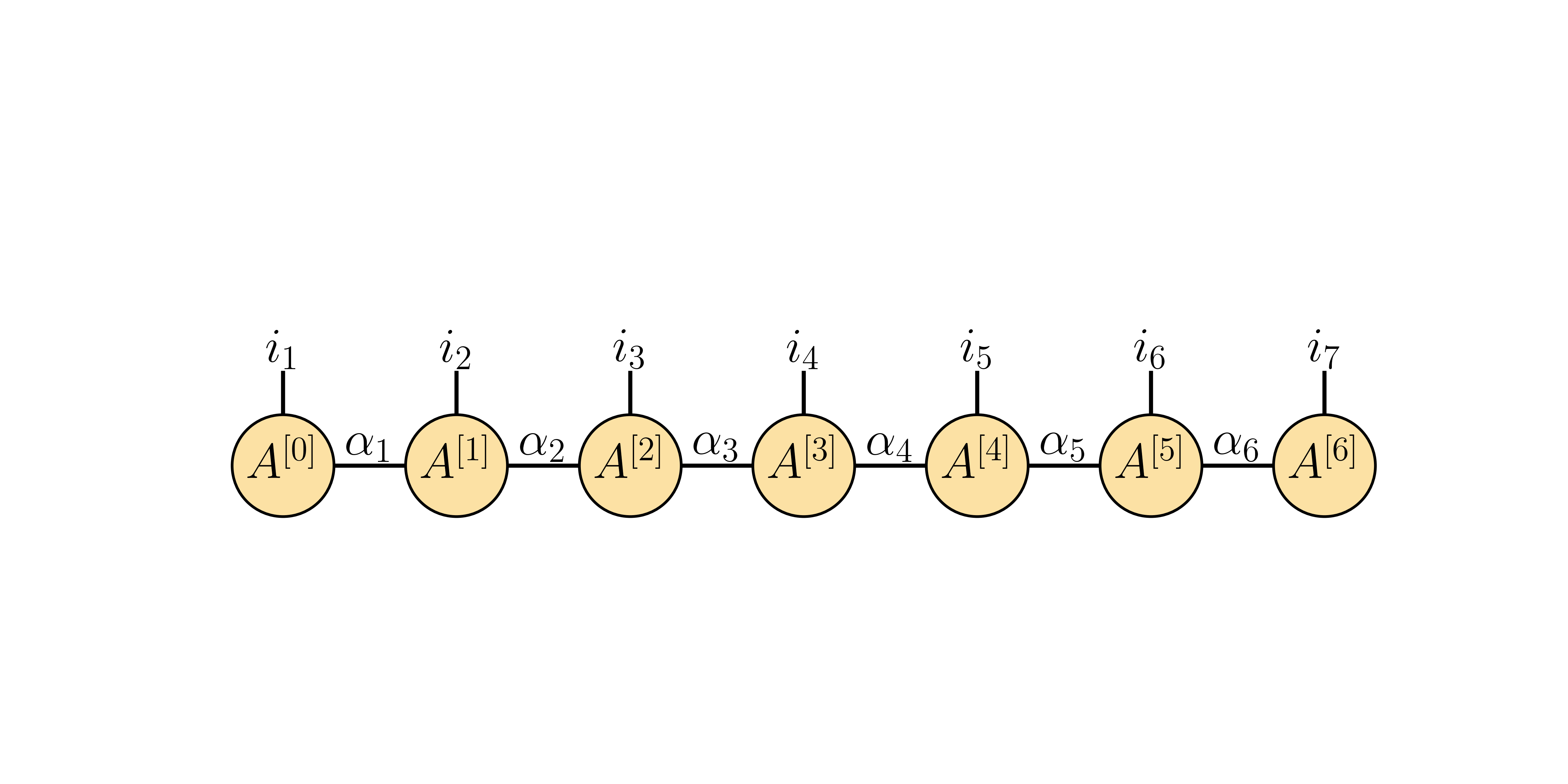}
    \caption{A visualisation of the structure of an open boundary condition MPS for an $n=7$ qubit state. The MPS cores $A^{[j]}$ have physical indices $i_j$ with dimension $d=2$ and are connected by virtual indices $\alpha_j$ with dimension at most $\chi$.}
    
    \label{fig:mps_vis}
\end{figure}

The matrix product state (MPS) or tensor train (TT) format of tensor networks is a factorisation of a tensor with $N$ indices into a chain-like product of three-index tensors~\cite{orus2014}. The MPS of an $n$-qubit system with open boundary conditions is given by:
\begin{equation}
    \ket{\psi} = \sum_{{i_1,...,i_n} = \{0,1\}} A^{[1]i_1}_{\alpha_1} A^{[2]i_2}_{\alpha_1,\alpha_2} ... A^{[n-1]i_{n-1}}_{\alpha_{n-2},\alpha_{n-1}} A^{[n]i_n}_{\alpha_{n-1}} \ket{i_1 i_2 ... i_n}
\label{eq:mps_definition}
\end{equation}
where $A^{[k]i_k}_{\alpha_{k-1},\alpha_{k}}$ is the MPS core of dimension $(\alpha_{k-1}$, 2, $\alpha_k)$ corresponding to the $k^{\text{th}}$ qubit, $\alpha_{k-1}$ is the left virtual index, $\alpha_{k}$ is the right virtual index, and $i_k$ is the physical index. The left and right virtual indices connect each MPS core to its neighbours, with the left virtual index of $A^{[0]}$ and the right virtual index of $A^{[n]}$ set to 1 by the convention of open boundary conditions \cite{Oseledets_tensor_train_decomposition}. The bond dimension of the MPS is defined as $\chi:=\max_{k}{\alpha_k}$ (that is, the bond dimension $\chi$ of an MPS is defined as the maximum of the dimensions of all virtual indices). An intuitive way to conceptualise the MPS construction is to imagine a one-dimensional spin-lattice consisting of $n$ sites, where the virtual indices carry information related to the entanglement between sites, as illustrated in Figure~\ref{fig:mps_vis}.

The open boundary MPS representation of a quantum state is non-unique. We can reduce the gauge redundancy in the representation by expressing the MPS in left-canonical form:
\begin{equation}
\begin{aligned}
    \sum_{i_1,\alpha_1} A^{[1]i_1}_{\alpha_1} (A^{[1]i_1}_{\alpha_1})^\dagger &= 1, \\
    \sum_{i_k,\alpha_k} A^{[k]i_k}_{\alpha_{k-1},\alpha_k} (A^{[k]i_k}_{\alpha_{k-1},\alpha_k})^\dagger &= \delta_{\alpha_{k-1}\alpha_{k-1}'}, \\
    \sum_{i_{n-1},\alpha_{n-1}} A^{[n]i_n}_{\alpha_{n-1}} (A^{[n]i_n}_{\alpha_{n-1}})^\dagger &= \delta_{\alpha_{n-1}\alpha_{n-1}'},
\end{aligned}
\label{eq:left_canonical_form}
\end{equation}
which defines the orthonormality constraints of the MPS.

\subsection*{3.2 Approximating Quantum States with MPS}

An $n$-qubit MPS of bond dimension $\chi$ contains $n$ tensors of dimension at most $(\chi,2,\chi)$. Hence, at most $2n\chi^2$ elements are required to specify the MPS. The exact specification of an arbitrary $n$-qubit state requires a bond dimension $\chi=\mathcal{O}(2^{n/2})$. We can optimally approximate states, however, using fixed-$\chi$ representations, improving the classical efficiency of the representation at the expense of an approximation error.

Consider a bipartition of a pure $n$-qubit quantum state $\ket{\Psi_{AB}}$ into subsystems $A$ and $B$ containing $j$ and $n-j$ sites, respectively. By reshaping $A^{[j-1]}$ and $A^{[j]}$ into a single matrix $M_{\mu\nu}$, we can compute the singular value decomposition (SVD) and truncate the smallest $s$ values. This truncates the virtual index $\alpha_j$ to $\alpha_j '=\alpha_j-s$. By the Eckart-Young-Mirsky theorem \cite{Eckart1936,mirsky}, this is an optimal approximation of the quantum state. The Frobenius norm error introduced by this truncation is given by
\begin{equation}
    \epsilon_{T,k}^{(F)} = \sqrt{\sum_{i=\alpha_j '+1}^{\alpha_j}\sigma_i^2}
    \label{eq:truncation_error}
\end{equation}
where $\sigma_i$ denotes the singular values. In other words, the truncation error is the $L^2$-norm of the $s$ singular values discarded. By iterating across all $n-1$ possible bipartitions of the state, we can form an optimal fixed-$\chi$ approximation of $\ket{\Psi_{AB}}$ satisfying $\forall j, \alpha_j'\leq \chi$.

The maximum entanglement entropy across two entropy $S$ across any possible bipartition of an $n$-qubit state represented by an MPS is bounded by
\begin{equation}
    S\leq\log_2{\chi}
    \label{eq:vne_bound}
\end{equation}
where $S$ is the von Neumann entropy across the bipartition and $\chi$ is the bond dimension of the MPS \cite{orus2014}. Equation~\ref{eq:vne_bound} bounds the entanglement entropy that a fixed-$\chi$ MPS can express. 

The MPS is classically efficient if $\chi\leq\mathcal{O}(\text{poly}(n))$, meaning that efficient MPS can precisely express states with entanglement entropy scaling poly-logarithmically in $n$. We are specifically interested in states that obey an area law of entanglement entropy, satisfying $S=\mathcal{O}(1)$ for one-dimensional states. Area-law states are characterised by entanglement entropy that grows with the boundary of the partition rather than the volume. This means that area-law states have an approximation error independent of $n$ such that $\chi=\mathcal{O}(1)$. Area-law states are characterised by exponentially decaying singular values, meaning that fixed-$\chi$ MPS approximations of such states can be efficient, accurate, and scalable \cite{orus2014}. 

\begin{algorithm}[!ht]
\caption{MPS From State Vector \protect{\cite{gundlapalli_deterministic_MPS_preparation}}}
\label{alg:mps_from_sv_alg}
\begin{algorithmic}[1]
    \REQUIRE $n$-qubit state $\ket{\psi}$, optional bond dimension $\chi$
    \ENSURE MPS tensors $\{A^{[1]}, \dots, A^{[n]}\}$ with bond dimension $\chi$
    
    \STATE Express $\ket{\psi} = \sum_{i_1,\dots,i_n} \psi(i_1,\dots,i_n)\ket{i_1,\dots,i_n}$
    \STATE Reshape $\psi$ to matrix $B(i_1; i_2,\dots,i_n)$ of size $2 \times 2^{n-1}$
    \STATE Perform SVD: $B = U \Sigma V^\dagger$
    \STATE Set $A^{[1]}_{i_1,\alpha_1} \gets U(i_1,\alpha_1)$
    \STATE Reshape $V^\dagger$ to $B(\alpha_1,i_2; i_3,\dots,i_n)$
    
    \FOR{$k = 2$ to $n-1$}
        \STATE SVD: $B = U \Sigma V^\dagger$
        \STATE Set $A^{[k]}_{\alpha_{k-1},i_k,\alpha_k} \gets U(\alpha_{k-1},i_k,\alpha_k)$
        \STATE Reshape $V^\dagger$ to $B(\alpha_k,i_{k+1}; i_{k+2},\dots,i_n)$
    \ENDFOR
    
    \STATE Set $A^{[n]}_{\alpha_{n-1},i_n} \gets B(\alpha_{n-1},i_n)$
    \IF{fixed bond dimension $\chi$}
        \STATE Truncate singular values to top $\chi$ at each SVD step
    \ENDIF
    \RETURN $\{A^{[1]}, \dots, A^{[n]}\}$
\end{algorithmic}
\end{algorithm}

\subsection*{3.3 Matrix Product Disentangler (MPD) Algorithm}

An MPS with $n$ sites and bond dimension $\chi$ can be exactly prepared with a single layer of (at most) $(\log_2{\chi}+1)$-qubit unitaries \cite{schon_sequential_MPS_generation,schon_sequential_MPS_generation_QED,ran}, i.e. elements of the unitary group $U(2\chi)$. Any $q$-qubit gate can be decomposed into a circuit of depth $\mathcal{O}(4^q)$ one- and two-qubit gates using the Quantum Shannon Decomposition \cite{shende_minimal_CNOT_decomposition_bound}. Using $q=\log_2{\chi}+1$, we have that the circuit depth of the exact MPS preparation scheme scales as $\mathcal{O}(n\chi_{\text{max}}^2)$ where big-O notation may hide significant scale factors in the decomposition process. Hence, we are motivated to adopt a more approximate MPS encoding scheme that avoids the polynomial dependence of circuit depth on $\chi$, such as the MPD algorithm.

The matrix product disentangler (MPD) algorithm introduced by Ran in 2020 \cite{ran} approximately maps an $n$ site MPS using $L$ sequential layers of 2-qubit $U(4)$ unitaries, as depicted in Figure~\ref{fig:ran_circuit}. Each layer is computed iteratively using an analytic method detailed in Algorithm 1. The layers are computed in reverse, beginning with the original MPS $\ket{\psi}=\ket{\psi_1}$ (i.e., $\ket{\psi_1}$ is initialised to the original MPS). At the $k^{\text{th}}$ iteration, the MPS $\ket{\psi_k}$ is approximated by a $\chi=2$ MPS denoted by $\ket{\tilde{\psi}_k}$. Then, the $k^{\text{th}}$ MPD unitary $\hat{U}_k$ is computed such that $\hat{U}_k\ket{\tilde{\psi}_k}\approx\ket{0}^{\otimes n}$, before the MPS is updated using $\ket{\psi_{k+1}}=\hat{U}_k\ket{\psi_k}$. Note that we can reduce computational costs by setting a maximum bond dimension $\chi_{\text{max}}$ for $\ket{\psi_k}$ for all $k=1,...,L$. Since $\chi_{\text{exact}}=\mathcal{O}(2^{n/2})$, setting $\chi_{\text{max}}<<\chi_{\text{exact}}$ is essential for large $n$. The unitary $\hat{U}_k$ is called the MPD unitary because it approximately disentangles $\ket{\psi_k}$ to $\ket{\psi_{k+1}}$. Each MPD unitary is a $\chi=2$ approximation, which can be exactly mapped into a single layer of $U(4)$ gates \cite{schon_sequential_MPS_generation}. After repeating the disentangling process until $k=L$, we are left with $L$ MPD unitaries that can each be collectively mapped onto $L$ layers of $U(4)$ gates similar to the circuit in Figure~\ref{fig:ran_circuit}. 

The MPD unitaries act such that $\hat{U}_1\hat{U}_2\cdots\hat{U}_L\ket{\psi}\approx\ket{0}^{\otimes n}$ where $\ket{\psi}$ is the target state. It follows that $\hat{U}^\dagger_L\hat{U}^\dagger_{L-1}\cdots\hat{U}^\dagger_1\ket{0}^{\otimes n}=\ket{\psi'}\approx\ket{\psi}$. Hence, we are interested in applying a sequence of inverse MPD unitaries $\{\hat{U}_k^\dagger\}_{k=1}^L$ to the ground state, which enables the target state to be approximately prepared. At iteration $k$, the quality of the MPD unitary depends on the $\chi=2$ approximation of $\ket{\psi_k}$, meaning that the algorithm's performance depends on these approximations. If the $\chi=2$ approximation is very poor, then the MPD unitary $\hat{U}_k$ barely disentangles $\ket{\psi_k}$ and the additional circuit layer produced by the MPD algorithm does not contribute to the quality of the prepared state. In the extreme worst case, the MPD algorithm prepares the $\chi=2$ approximation of $\ket{\psi}$, and subsequent layers do not contribute. This is consistent with the findings in \cite{BenDov2024} that the MPD approach exhibits diminishing gains in fidelity with increasing layers, and it is not evident that the state preparation error $\epsilon_S\rightarrow 0$ as $L\rightarrow\infty$. The quality of the $\chi=2$ approximations corresponds to the decay rate of the singular values in the MPD, meaning that the MPD algorithm will perform best in preparing states with rapidly decaying singular values. A key pitfall, however, is that no rigorous performance guarantees exist, which motivates the extensive numerical study in this article.

\begin{figure}[!ht]
\[
\Qcircuit @C=0.55em @R=0.75em {
    \lstick{\ket{0}} & 
    \qw                             & \qw                             & \qw                             & \qw                             & 
    \multigate{1}{\hat{U}_{1}^{\dagger (5)}} & \gate{\hat{U}_{1}^{\dagger (6)}} & \qw                             & \qw                             & \qw                             & \qw                             & 
    \multigate{5}{\ \hat{U}_2^\dagger\ }     & \qw & \push{\cdots} & & \qw & \multigate{5}{\ \hat{U}_L^\dagger\ } & \meter \\
    \lstick{\ket{0}} & 
    \qw                             & \qw                             & \qw                             & 
    \multigate{1}{\hat{U}_{1}^{\dagger (4)}} & \ghost{\hat{U}_{1}^{\dagger (5)}} & \qw                             & \qw                             & \qw                             & \qw                             & \qw                             &
    \ghost{\ \hat{U}_2^\dagger\ }            & \qw & \push{\cdots} & & \qw & \ghost{\ \hat{U}_L^\dagger\ }         & \meter \\
    \lstick{\ket{0}} & 
    \qw                             & \qw                             & \multigate{1}{\hat{U}_{1}^{\dagger (3)}} & \ghost{\hat{U}_{1}^{\dagger (4)}} & \qw                             & \qw                             & \qw                             & \qw                             & \qw                             & \qw                             &
    \ghost{\ \hat{U}_2^\dagger\ }            & \qw & \push{\cdots} & & \qw & \ghost{\ \hat{U}_L^\dagger\ }         & \meter \\
    \lstick{\ket{0}} & 
    \qw                             & \multigate{1}{\hat{U}_{1}^{\dagger (2)}} & \ghost{\hat{U}_{1}^{\dagger (3)}} & \qw                             & \qw                             & \qw                             & \qw                             & \qw                             & \qw                             & \qw                             &
    \ghost{\ \hat{U}_2^\dagger\ }            & \qw & \push{\cdots} & & \qw & \ghost{\ \hat{U}_L^\dagger\ }         & \meter \\
    \lstick{\ket{0}} & 
    \multigate{1}{\hat{U}_{1}^{\dagger (1)}} & \ghost{\hat{U}_{1}^{\dagger (2)}} & \qw                             & \qw                             & \qw                             & \qw                             & \qw                             & \qw                             & \qw                             & \qw                             &
    \ghost{\ \hat{U}_2^\dagger\ }            & \qw & \push{\cdots} & & \qw & \ghost{\ \hat{U}_L^\dagger\ }         & \meter \\
    \lstick{\ket{0}} & 
    \ghost{\hat{U}_{1}^{\dagger (1)}}        & \qw                             & \qw                             & \qw                             & \qw                             & \qw                             & \qw                             & \qw                             & \qw                             & \qw                             &
    \ghost{\ \hat{U}_2^\dagger\ }            & \qw & \push{\cdots} & & \qw & \ghost{\ \hat{U}_L^\dagger\ }         & \meter \gategroup{1}{2}{6}{7}{.7em}{--} \\
}
\]
        \captionsetup{justification=centering, singlelinecheck=off}
        \caption{The sequential circuit for the MPD and MPD+TNO algorithms. Each entangling layer $\hat{U}_j^\dagger$ for $j=1,2,...,L$ corresponds to a $\chi=2$ MPO as computed via Algorithm 2. The first circuit layer $\hat{U}_1^\dagger$ exactly prepares the $\chi=2$ approximation of the target state, with subsequent layers designed to progressively increase entanglement and improve upon the $\chi=2$ approximation. The maximum number of layers $L$ is a hyperparameter.}
    \label{fig:ran_circuit}
\end{figure}

\begin{algorithm}[!ht]
\caption{Analytic Matrix Product Disentangler (MPD) Algorithm \cite{ran}}
\label{alg:ran_mps}
\begin{algorithmic}[1]
    \REQUIRE Target MPS $\ket{\psi_1}$ with bond dimension $\chi_{\text{max}}$, number of layers $L$
    \ENSURE Set of unitary layers $\{\hat{U}^\dagger_k\}_{k=1}^{L}$
    \STATE Initialize $k \gets 1$
    \WHILE{$k \leq L$}
        \STATE Approximate $\ket{\psi_k}$ by $\ket{\tilde{\psi}_k}$ with bond dimension $\tilde{\chi}=2$
        \STATE Compute the MPD $\hat{U}_k$ such that $\hat{U}_k \ket{\tilde{\psi}_k} \approx \ket{0}^{\otimes n}$
        \STATE Approximately disentangle $\ket{\psi_k}$ to $\ket{\psi_{k+1}}$ using $\ket{\psi_{k+1}} = \hat{U}_k \ket{\psi_k}$
        \STATE Update $k \gets k + 1$
    \ENDWHILE
    \RETURN The set $\{\hat{U}^\dagger_k\}_{k=1}^{L-}$
    \STATE \textbf{Note:} The circuit $U_S := \hat{U}_{L}^\dagger \hat{U}_{L-1}^\dagger \cdots \hat{U}_{1}^\dagger$ applied to the zero state $\ket{0}^{\otimes n}$ approximately reconstructs the input MPS $\ket{\psi_1}$
\end{algorithmic}
\end{algorithm}

\subsection*{3.4 Tensor Network Optimisation (TNO)}

We aim to investigate the efficacy and applications of the MPD algorithm combined with Tensor Network Optimisation (TNO), which we denote as the MPD+TNO algorithm. To begin with, we compute the MPD algorithm with $L$ circuit layers. The output of the MPD algorithm is a quantum circuit consisting of $L$ layers of $U(4)$ unitaries (see Figure~\ref{fig:ran_circuit}) that prepare an approximation of the target state. This quantum circuit is decomposed into arbitrary single-qubit rotation gates and CNOT gates. Each single-qubit rotation is parameterised by 3 Euler angles, and the total number of parameters in the circuit scales as $\mathcal{O}(nL)$ (with $n$ being the number of qubits). The parameterised quantum circuit is represented as a tensor network to facilitate efficient contraction and differentiation.

The cost function is defined as,
\begin{equation}
    L(\theta) = 1 - |\braket{\psi_{\text{mps}}|\psi_{\text{circuit}}(\theta)}|^2
\end{equation}
where $\psi_{\text{mps}}$ is the target state represented as an MPS with bond dimension $\chi_{\text{max}}$ and expressed in left-canonical form, and $\psi_{\text{circuit}}(\theta)$ is the parameterised quantum circuit expressed as a tensor network. We can compute this cost function efficiently by leveraging efficient tensor contractions between the parameterised circuit and the target MPS (i.e., it circumvents the exponential cost associated with full state-vector simulation). 

Gradient descent is computed via the widely employed Limited-memory Broyden-Fletcher-Goldfarb-Shanno (L-BFGS-B), which is a quasi-Newton method that approximates the Hessian matrix $H$ (the second derivative of $L(\theta)$ w.r.t. $\theta$). The parameters at iteration $t+1$ are updated as follows,
\begin{equation}
    \theta^{(t=1)} = \theta^{(t)} - \eta H_t\nabla L(\theta^{(t)})
\end{equation}
where $\theta$ is the set of parameters, $\eta$ is the learning rate, and $H_t$ is the approximate Hessian matrix at iteration $t$. The gradients are computed with automatic differentiation. 

The efficiency of this scheme depends on the degree of initial overlap $|\braket{\psi_{\text{mps}}|\psi_{\text{MPD circuit}}}|$. This can be thought of as initialising the parameters near the global minimum, reducing the number of iterations required to converge and avoiding suboptimal local minima \cite{grant2019initialization_barrenplateaus}. The resulting quantum circuit can be retrieved via efficient contractions of the tensor network \cite{orus2014,gray_hyper_optimised_tn_contraction}.  Potentially, alternative approaches to gradient descent and optimisation strategy could be considered. For instance, it was shown by Rudolph
et al. \cite{rudolph2022decompositionmatrixproductstates} that iteratively optimising each additional layer (rather than all the layers at once) can improve training efficiency in some instances. This alternative approach comes with a greater (quadratic) dependence of complexity on the number of iterations $T$ scaling as $\mathcal{O}(nT^2L\chi^3)$. 

An alternative TNO strategy developed by Melnikov et al. \cite{melnikov} was also considered. This scheme employs the hardware efficient ansatz (HEA) consisting of layers of single-qubit rotations and CNOT gates, optimised via Riemannian optimisation over the Stiefel manifold. This approach initialises the circuit parameters by training subsets of the quantum circuit represented as MPSs, before recombining them and training on the full quantum circuit. However, this algorithm's performance was unsatisfactory, owing to the relatively low overlap with the global minimum obtained from this initialisation strategy. In contrast, using the MPD algorithm to initialise parameters before TNO leads to significantly higher overlap between the initial state and the global minimum, thereby increasing training efficiency. This illustrates the importance of high-quality parameter initialisation for training efficiency in MPS-based optimisation strategies.

\subsection*{3.5 Complexity Analysis}

We consider the complexity analysis of the MPD and MPD+TNO methods, where $n$ denotes the number of qubits, $L$ denotes the number of layers, $T$ denotes the number of optimisation iterations, and $\chi_{\text{max}}$ denotes the maximum bond dimension used to approximate the target state. We consider the complexity of the following components of each algorithm:
\begin{itemize}
    \item The memory requirements of the MPS scale as $\mathcal{O}(n\chi_{\text{max}}^2)$. The cost of bringing the MPS into canonical form is $\mathcal{O}(n\chi_{\text{max}}^3)$ \cite{Oseledets_tensor_train_decomposition}. The cost of the bond dimension truncation step used in the MPD algorithm scales as $\mathcal{O}(\chi_{\text{max}}^3)$.
    \item Computing the inner product between $\ket{\psi_{\text{mps}}}$ and $\ket{\psi_{\text{circuit}}}$ scales as $\mathcal{O}(n\chi_{\text{max}}^3)$ if both MPS have bond dimension $\chi_{\text{max}}$.
    \item It follows that the MPD algorithm scales as $\mathcal{O}(nL\chi_{\text{max}}^3)$ and the MPD+TNO algorithm as $\mathcal{O}(nL\chi_{\text{max}}^3*T)$ where $T$ is the number of iterations required to converge to a solution. If we use a layer-by-layer optimisation, as introduced in \cite{rudolph2022decompositionmatrixproductstates}, the training complexity becomes $\mathcal{O}(nL\chi_{\text{max}}^3*T^2)$
    \item We must also consider the cost of decomposing arbitrary $U(q)$ unitaries into elementary gates. The Quantum Shannon Decomposition approach to this problem encompasses a $\mathcal{O}(8^q)$ classical overhead and a circuit depth $\mathcal{O}(4^q)$ for each $q$-qubit gate \cite{krol_unitary_matrix_decomposition}. In the special case of $q=2$, any $U(4)$ unitary can be decomposed into \textit{at most} 3 CNOT gates and 15 elementary single-qubit operations, which is optimal \cite{optimal_2_qubit_gate_decomp}. Hence, the big-O notation in the MPD and MPD+TNO algorithms hides only modest scale factors. Accounting for elementary gate decomposition into single-qubit rotations and CNOT gates, the elementary circuit depth in the exact method of \cite{schon_sequential_MPS_generation} is $\mathcal{O}(n\cdot4^{\log_2{(\chi_{\text{max}}+1)}}) = \mathcal{O}(n\cdot4\chi_{\text{max}}^2) = \mathcal{O}(n\chi_{\text{max}}^2)$.
\end{itemize}

\begin{table}[!htbp]
\centering
\small
\setlength{\tabcolsep}{4pt}
\renewcommand{\arraystretch}{1.15}
\setlength{\extrarowheight}{2.0pt} 

\resizebox{\textwidth}{!}{%
\begin{tabular}{|c|c|c|c|c|c|}
\hline
\parbox[c][2.0\baselineskip][c]{3cm}{\centering\textbf{Method}} &
\parbox[c][2.0\baselineskip][c]{3cm}{\centering\textbf{Circuit Type}} &
\parbox[c][2.0\baselineskip][c]{2.5cm}{\centering\textbf{Time}} &
\parbox[c][2.0\baselineskip][c]{2cm}{\centering\textbf{Depth}} &
\parbox[c][2.0\baselineskip][c]{2cm}{\centering\textbf{Ancillas}} &
\parbox[c][2.0\baselineskip][c]{4cm}{\centering\textbf{Comments}} \\
\hline

\parbox[c]{3cm}{\centering MPD \cite{ran}} &
\parbox[c]{3cm}{\centering $L$ Layers of $U(4)$\\Sequential} &
\parbox[c]{2.5cm}{\centering $\mathcal{O}(nL\chi_{\text{max}}^3)$} &
\parbox[c]{2cm}{\centering $\mathcal{O}(nL)$} &
\parbox[c]{2cm}{\centering 0} &
\parbox[c]{4cm}{\centering Modest hidden scale factors} \\
\hline

\parbox[c]{3cm}{\centering Melnikov et al. TNO \cite{melnikov}} &
\parbox[c]{3cm}{\centering $L$ Layers of\\Hardware Efficient Ansatz} &
\parbox[c]{2.5cm}{\centering $\mathcal{O}(nTL\chi_{\text{max}}^3)$} &
\parbox[c]{2cm}{\centering $\mathcal{O}(nL)$} &
\parbox[c]{2cm}{\centering 0} &
\parbox[c]{4cm}{\centering MPS subsets trained to initialise parameters} \\
\hline

\parbox[c]{3cm}{\centering MPD+TNO \cite{rudolph2022decompositionmatrixproductstates,BenDov2024}} &
\parbox[c]{3cm}{\centering $L$ Layers of $U(4)$\\Sequential} &
\parbox[c]{2.5cm}{\centering $\mathcal{O}(nTL\chi_{\text{max}}^3)$} &
\parbox[c]{2cm}{\centering $\mathcal{O}(nL)$} &
\parbox[c]{2cm}{\centering 0} &
\parbox[c]{4cm}{\centering MPD algorithm used to initialise parameters} \\
\hline

\parbox[c]{3cm}{\centering MPD+Layer-by-Layer TNO \cite{rudolph2022decompositionmatrixproductstates}} &
\parbox[c]{3cm}{\centering $L$ Layers of $U(4)$\\Sequential} &
\parbox[c]{2.5cm}{\centering $\mathcal{O}(nT^2L\chi_{\text{max}}^3)$} &
\parbox[c]{2cm}{\centering $\mathcal{O}(nL)$} &
\parbox[c]{2cm}{\centering 0} &
\parbox[c]{4cm}{\centering Full MPS optimised with each new layer} \\
\hline
\end{tabular}%
}

\caption{A comparison of the complexity of MPS encoding schemes.}
\label{tab:t7}
\end{table}

\section*{4. MPS Preparation of Functions}

This section explores the viability of the MPD and MPD+TNO algorithms for preparing states representing a vast array of discretised elementary and polynomial functions. We ultimately find the MPS approach exceptionally efficient in preparing high-fidelity approximations of functions up to low-degree piecewise polynomials with linear-depth circuits and efficient classical pre-processing costs. 

\subsection*{4.1 Previous Work and Gaps}
MPS encoding has been identified as a highly NISQ-friendly framework for state preparation \cite{ran,BenDov2024,malz_mps_encoding,gonzalez,rudolph2022decompositionmatrixproductstates,iqbal_2022_mps}. Holmes \& Matsuura (2020) \cite{holmes_smooth_diff_functions}, motivated by the bound on entanglement entropy for discretised functions \cite{Garc_a_Ripoll_2021_quantuminspired}, explored applications of the MPS framework for the state preparation of smooth, differentiable functions. In particular, \cite{holmes_smooth_diff_functions} proposed the computation of the MPS corresponding to a piecewise polynomial approximation of the target function before the MPD algorithm is used to construct the quantum circuit. \cite{holmes_smooth_diff_functions} ultimately demonstrated the efficacy of this scheme for encoding $\chi=2$ approximations of the Gaussian, Lognormal, and Lorentzian family of functions. Additionally, Iaconis et al. (2024) \cite{iaconis} employed the MPD algorithm to encode Gaussian functions in the \textit{probabilities} of the state. These approaches, however, did not optimise the prepared quantum circuits, and it remains unclear how well these results can generalise to functions exhibiting more irregular or complex behaviour.

Recently, TNO methods for MPS preparation have emerged as viable encoding schemes \cite{BenDov2024,rudolph2022decompositionmatrixproductstates,melnikov,lubasch_nonlinear_vqa}. The authors in \cite{melnikov,lubasch_modular_QMC_integration}, for example, employ a TNO scheme that optimises state preparation circuits over the brick-wall structure, though these methods lack an efficient parameter initialisation. In \cite{BenDov2024,rudolph2022decompositionmatrixproductstates}, the MPD algorithm is used to initialise the parameters before TNO techniques are applied, which is the MPD+TNO methodology we explore in this paper. However, research into the applications and usefulness of this approach in practical problems like function encoding has been limited.

MPS of constant bond dimension can also be exactly prepared with a unitary circuit of depth $\mathcal{O}(n)$ \cite{schon_sequential_MPS_generation,schon_sequential_MPS_generation_QED,perezgarcia2007matrixproductstaterepresentations}. This follows the exact decomposition method of Sch\"on (2005) \cite{schon_sequential_MPS_generation}, which maps the MPS into a circuit of arbitrary ($\log_2{\chi}+1)$-qubit unitaries. To be precise, each virtual index $\alpha$ can be mapped to an ($\log_2{\alpha}+1)$-qubit unitary, which forms a single-layer sequential circuit similar to Figure~\ref{fig:ran_circuit} except with larger multi-qubit unitaries. The theoretical lower-bound on the unitary decomposition of a $q$-qubit unitary into one- and two-qubit gates is $\Omega(\frac{1}{4}4^q)$ CNOT gates \cite{shende_minimal_CNOT_decomposition_bound}, and the Quantum Shannon Decomposition can achieve a scaling of $\mathcal{O}(4^q)$ circuit depth \cite{krol_unitary_matrix_decomposition}. This implies that the decomposition of the circuit in \cite{schon_sequential_MPS_generation} results in a depth of $\mathcal{O}(n\cdot4^{\log_2{\chi+1}})=\mathcal{O}(4n\chi^2)$. However, the circuit decomposition can come with a significant overhead, which motivates the adoption of an approximate MPS encoding scheme like the MPD algorithm. Nonetheless, this implies that a circuit of depth $\mathcal{O}(4n\chi^2)$ can exactly encode the MPS. 

\subsection*{4.2 MPS Representations of Discretised Functions}

\subsubsection*{States Representing Discretised Functions}

We consider target states of the form given by,
\begin{equation}
    \ket{\psi} = \frac{1}{\|f\|_2} \sum_{j=0}^{N-1} f(x_j) \ket{j}
    \label{eq:target_state_func}
\end{equation}
where $f: \mathbb{R}\rightarrow\mathbb{R}$ is a univariate real-valued function defined over the domain $D=[a,b]$, $x_j = a + j\frac{b-a}{N-1}$ for a uniform discretisation, $N=2^n$ is the dimension of the $n$-qubit state vector $\ket{\psi}$, and $||f||_2$ is the Euclidean norm.

We first investigate the capabilities of MPS structures to represent functions with small bond dimensions accurately. The limiting factor in MPS state preparation is the bond dimension that can be prepared by the approximate quantum circuit encoding. Hence, we must ask what states can be represented by small bond dimensions without significant errors. We measure this by the truncation error, defined as:
\begin{equation}
    \epsilon_T = 1 -|\braket{\psi|\tilde{\psi}}|^2
\end{equation}
where $\ket{\psi}$ represents the exact target state and $\ket{\tilde{\psi}}$ represents the approximated target state achieved by optimally truncating the bond dimension to $\chi<\chi_{\text{exact}}$. In other words, the truncation error is the infidelity of the optimally approximated target state of a specified bond dimension $\chi$. 

The discretisation error of a real-valued and smooth function across a uniform grid of $2^n$ points in the domain $D =[a,b] \subset \mathbb{R}$ scales as $\mathcal{O}(2^{-n})$. Moreover, the maximum amount of von Neumann entropy (across any bipartition of the quantum system) that can be introduced when adding one qubit to a state representing a real-valued and smooth discrete function is bounded by:
    \begin{equation}
    \Delta S \leq \frac{(b-a) \sqrt{||f'||_\infty}}{2^{n/2-1}} = \mathcal{O}(2^{-n/2})
    \label{eq:vne}
    \end{equation}
where $||f'||_\infty=\max_{x\in[a,b]}{|f'(x)|}$ and $f(x)$ is discretised over the domain $[a,b]$ \cite{Garc_a_Ripoll_2021_quantuminspired}. Assuming that $f(x)$ is Lipschitz continuous (it has bounded derivatives almost everywhere), then the upper bound is dominated by the exponential decay with system size $n$. This implies that functions representative of discrete functions generally have limited entanglement and weak correlations, which typically grow with the maximum derivative of the function in the relevant domain. Consequently, we expect many discrete functions to have limited entanglement entropy, making them suitable for efficient and accurate MPS representations. Additionally, consider fitting a polynomial to an arbitrary, random state vector. The maximum derivative of such a function would grow unbounded as qubits are added to the system, allowing the state to accumulate a near maximum amount of entanglement entropy and exhibit long-range solid correlations. Hence, this framework quantifies the transition from limited entanglement entropy in straightforward discrete functions to large entanglement entropy expected of random, arbitrary state vectors.

\subsubsection*{Bond Dimension Considerations}

There is an explicit MPS construction for polynomials $f(x)=\sum_{j=0}^d a_jx^j$ of degree $d$ with a bounded $\chi\leq d+1$ \cite{oseledets2013constructive_mps_functions}. \cite{holmes_smooth_diff_functions} extended this construction to degree-$d$ piecewise polynomials defined across $2^I$ subdomains (i.e. $I$ piecewise intervals), which accumulates a maximum bond dimension of $\chi=2^I(d+1)$. These provide upper bounds on the exact $\chi$ required to represent these states. By considering the exact polynomial required to represent a given function, we can determine the maximum bond dimension to express it precisely. However, in many cases, a function can be well-approximated by a polynomial of a much lower degree. Additionally, the MPS of functions with periodic behaviour, like sinusoidal functions, can permit near-exact MPS representations with low bond dimensions \cite{oseledets2013constructive_mps_functions}. The transition in the fidelity of the prepared state with increasing bond dimension can be visualised in the simple example of a Gaussian function in Figure~\ref{fig:bond_dimension_transition}.

\begin{figure}[H]
    \centering
    \includegraphics[trim=0cm 0cm 0cm 0cm, clip, width=\textwidth]{MPS_Truncation_Error_Analysis/bond_dimension_transition.png}
    \caption{Visual improvement in the fidelity of the  n = 12-site MPS representation of a discretised Gaussian function $f(x)\sim N(0,0.3^2)$ with increasing bond dimension $\chi$.
    }
    \label{fig:bond_dimension_transition}
\end{figure}

\subsubsection*{Elementary Functions}

\begin{figure}[!htbp]
    \centering
    \begin{subfigure}{0.5\textwidth}
        \centering
        \includegraphics[width=\textwidth]{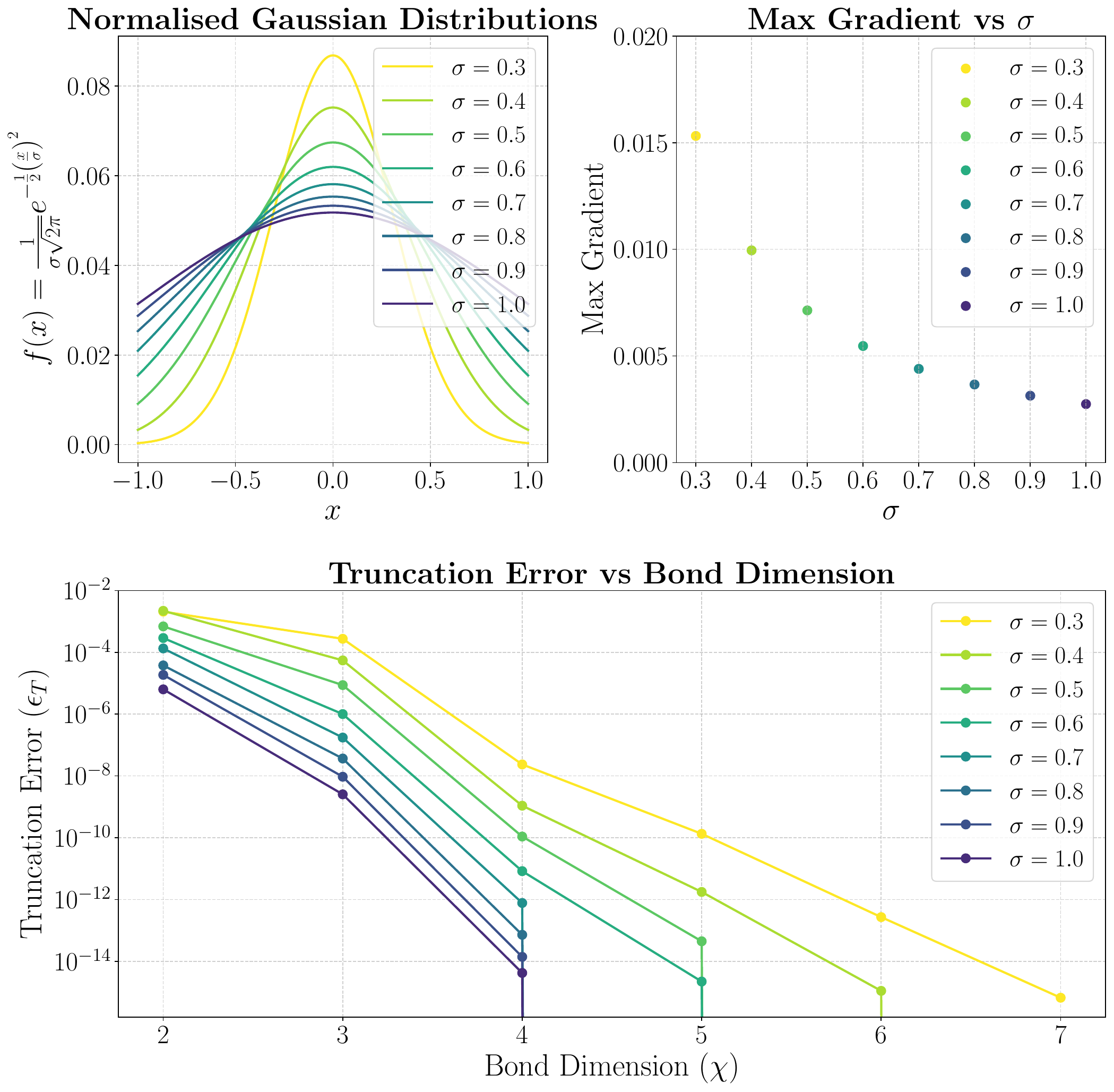} 
        \caption{}
        \label{fig:big}
    \end{subfigure}%
    \hfill
    \begin{subfigure}{0.49\textwidth}
        \centering
        \begin{subfigure}{\textwidth}
            \centering
            \includegraphics[width=\textwidth]{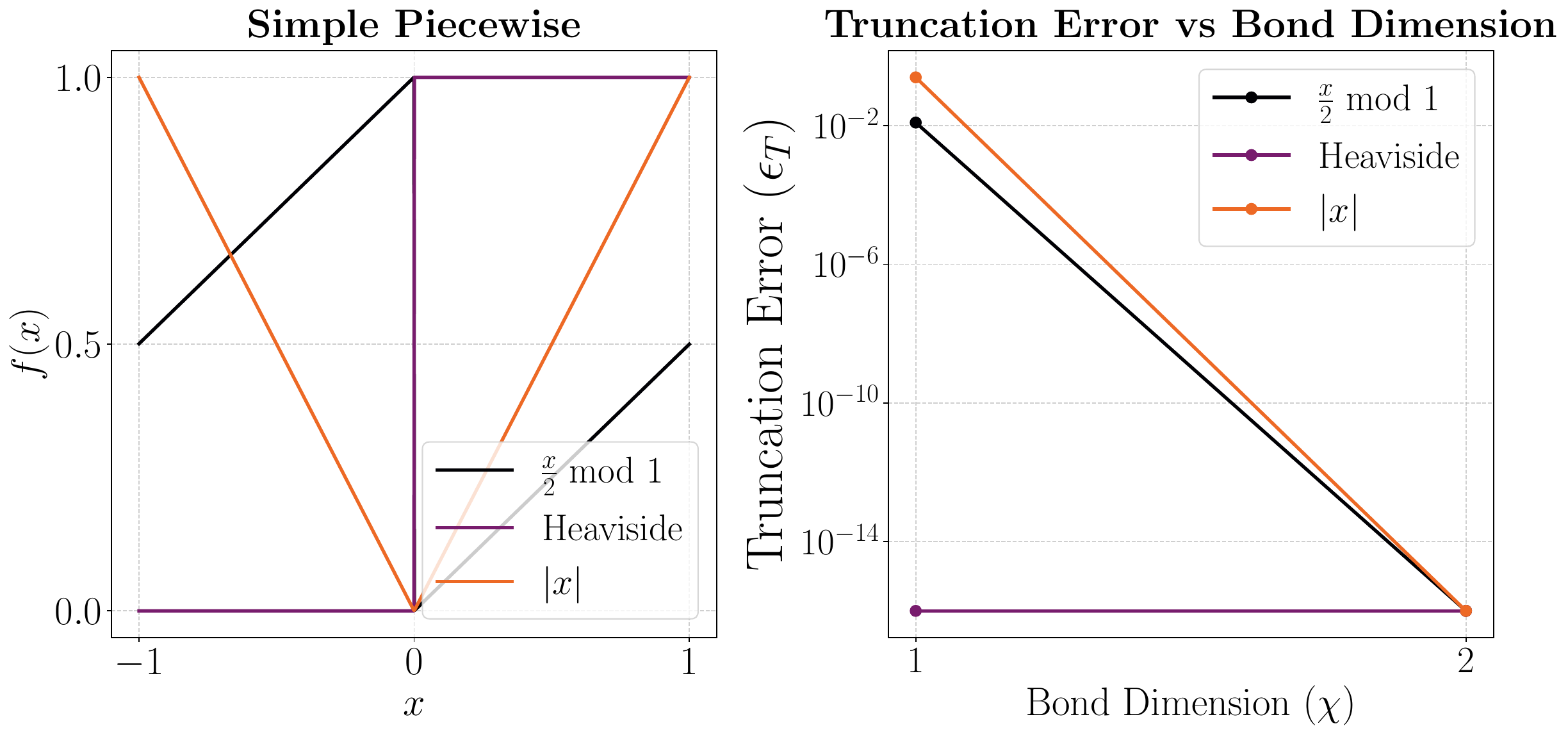}
            \caption{}
            \label{fig:small1}
        \end{subfigure}
        \vspace{0cm} 
        \begin{subfigure}{\textwidth}
            \centering
            \includegraphics[width=\textwidth]{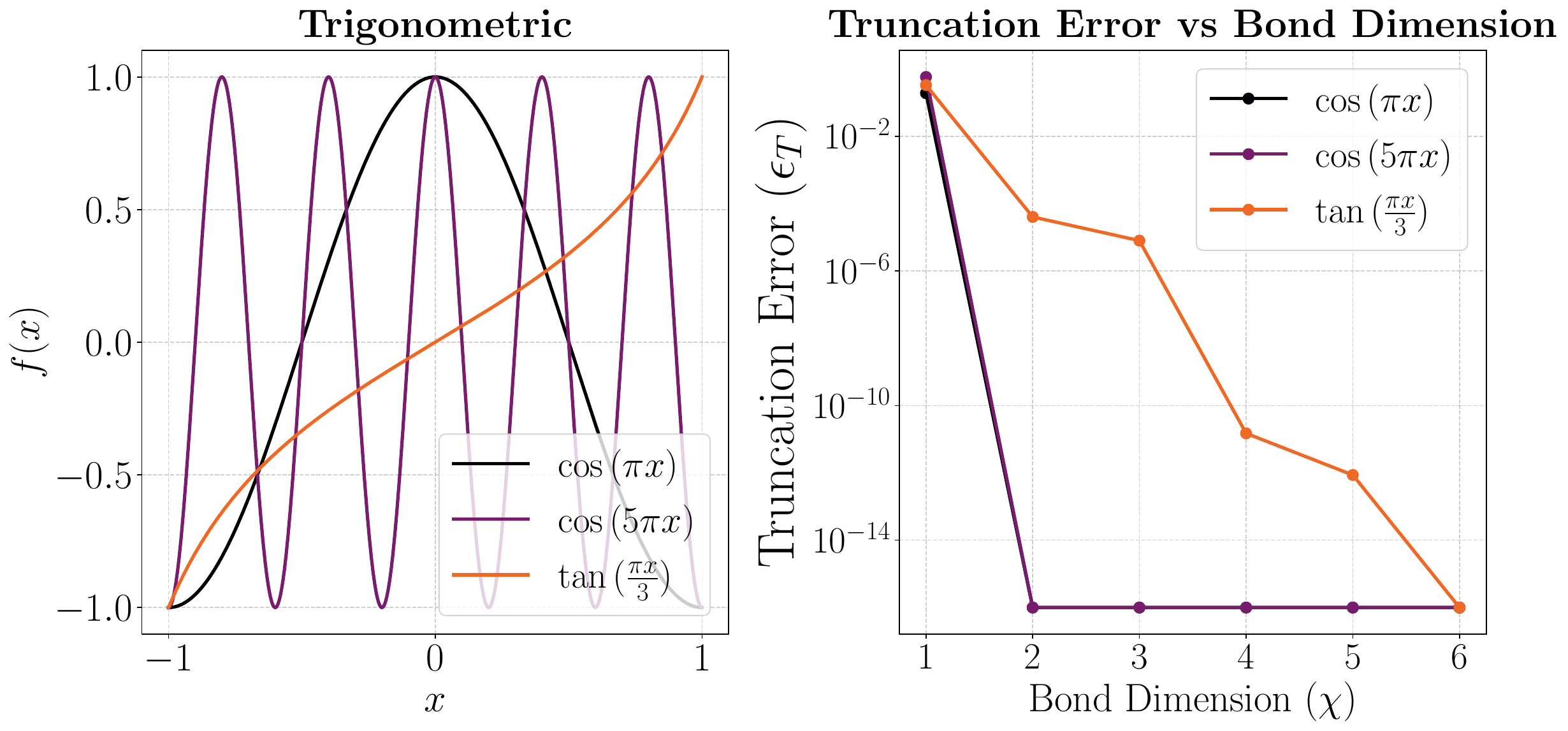}
            \caption{}
            \label{fig:small2}
        \end{subfigure}
    \end{subfigure}

    \vspace{0cm} 

    \begin{subfigure}{0.49\textwidth}
        \centering
        \includegraphics[width=\textwidth]{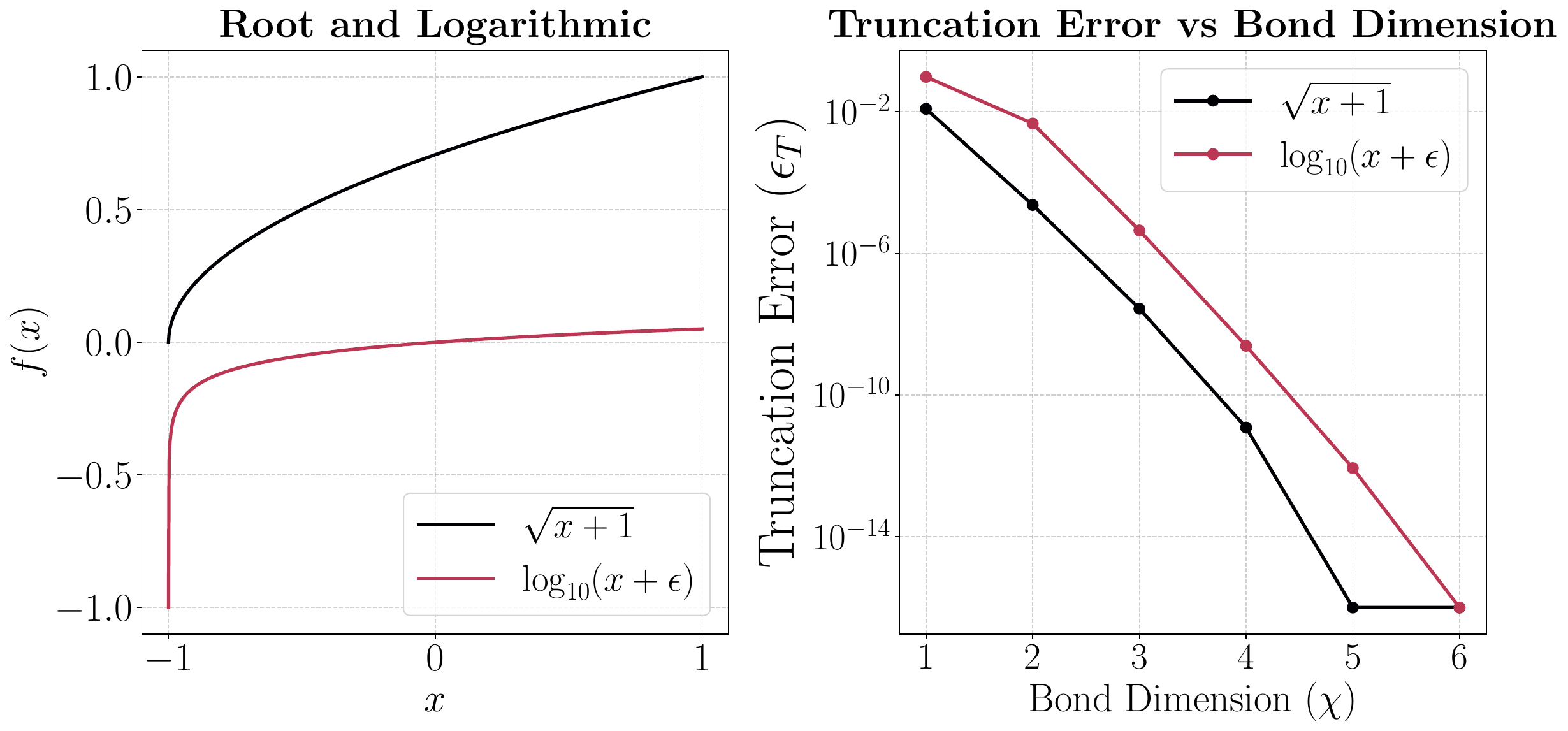} 
        \caption{}
        \label{fig:small3}
    \end{subfigure}%
    \hfill
    \begin{subfigure}{0.49\textwidth}
        \centering
        \includegraphics[width=\textwidth]{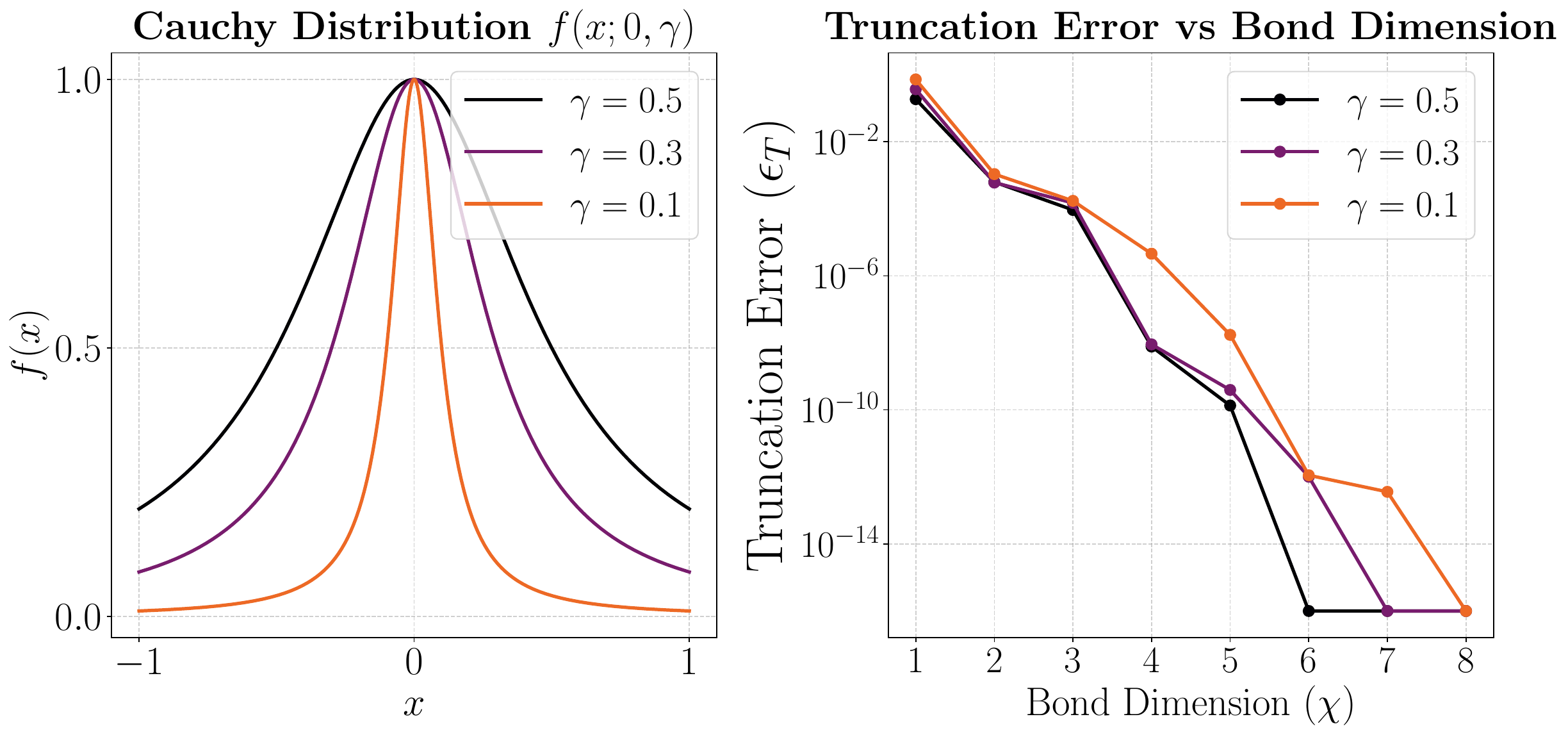} 
        \caption{}
        \label{fig:small4}
    \end{subfigure}

    \caption{Truncation Error Analysis for Fixed-$\chi$ Elementary Functions. The truncation error for $n=16$-qubit fixed-$\chi$ MPS representations is computed using Algorithm 1. The range of elementary functions explored includes (a) The Gaussian family, (b) Simple, linear piecewise functions, (c) Basic trigonometric functions, (d) Root and logarithmic functions, and (e) The Cauchy family. 
    }
    \label{fig:elementary_mps_truncation_error}
\end{figure}

First, we investigated the approximate representation of discretised elementary functions, such as straightforward probability distributions. We began by reproducing the findings of \cite{holmes_smooth_diff_functions} that MPS structures can efficiently represent Gaussian functions $f(x)\sim N(0,\sigma^2)$. Rather than assessing the $\chi=2$ representation of piecewise polynomials (fit to Gaussian functions), as is done in \cite{holmes_smooth_diff_functions}, this study delineated the \textit{exact} truncation error ($\epsilon_T$) of a fixed-$\chi$ MPS across a range of $\chi$ values. Results for Gaussian functions of varying $\sigma$ are presented in Figure~\ref{fig:elementary_mps_truncation_error}(a). The truncation error is universally larger at each $\chi$ as $\sigma$ decreases. This aligned with the expectation that the increased maximum gradient in these functions raises the bound on entanglement entropy scaling in Equation~\ref{eq:vne}. This effect was referred to as a ``squeezing" of the function in \cite{holmes_smooth_diff_functions}. By Equation~\ref{eq:vne_bound}, the bond dimension required to capture the increased entanglement entropy must also increase. Equivalently, the singular values in the SVD across any bipartition decay more slowly, resulting in a larger error when truncating the bond dimension.

Figure~\ref{fig:elementary_mps_truncation_error}(b)-(e) extended this work to a myriad of discretised functions defined on the domain $D=[-1,1]$. This included simple piecewise linear functions, such as the Heaviside step function and $f(x)=|x|$, basic trigonometric functions, the root function $f(x)=\sqrt{x+1}$, the logarithmic function $f(x)=\log_{10}{(x+1+\epsilon)}$ ($\epsilon=1e-6$), and the Cauchy distribution for varying $\gamma$ parameters. The Heaviside step function is a separable state with an exact $\chi=1$ representation. The linear and sinusoidal functions were demonstrated to have an exact $\chi=2$ representation, which aligns with the exact bond dimension of the \textit{explicit} MPS construction of these states provided in \cite{oseledets2013constructive_mps_functions}. Functions exhibiting increasingly irregular behaviour, with sharp gradients and asymmetries, were anticipated to necessitate a larger bond dimension $\chi$. Nonetheless, all explored functions have a near-exact MPS representation with a bond dimension at most $\chi\sim 8$, owing to the predominantly smooth behaviour of the functions. This is despite the significant discontinuity in the modulo function $f(x)=\frac{x}{2} \mod 1$ and the singularities at $x=-1$ in the root and logarithmic functions, illustrating that these states have bounded entanglement entropy despite their irregular characteristics. It was also demonstrated that high-quality approximations of each state correspond to a much lower-than-exact bond dimension, exchanging increased efficiency for decreased accuracy in the MPS representation. This is compelling evidence that the MPS structures are adept at accurately representing quantum states representative of discretised functions, exploiting limited entanglement entropy to overcome the dimensionality burden through an efficiently compressed representation of the states.

The truncation error at varying `singular value thresholds' refers to truncating all singular values below a minimum threshold to approximate the MPS. All truncation error results are computed for $n=16$ qubits. However, the singular value decays are computed for $n=8$ to help visualise the varying singular value decay rates. These results help to establish a visual intuition behind the capability of efficient MPS to accurately encode functions being dependent on the decay rate of the singular values in the SVD across each virtual index.

\subsubsection*{Polynomial Functions}

\begin{figure}[htbp]
    \centering

    \begin{subfigure}{0.49\textwidth}
        \centering
        \includegraphics[width=\textwidth]{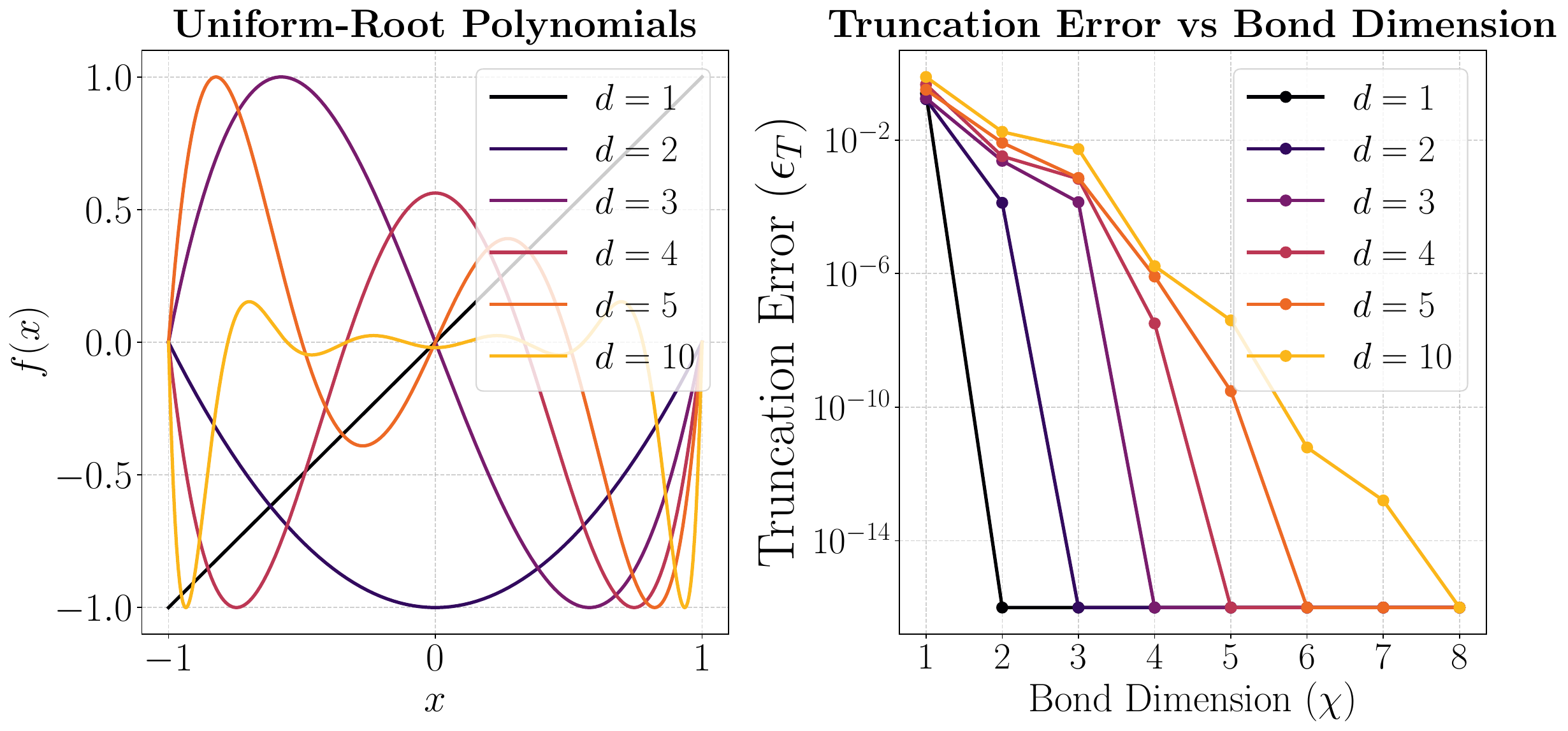}
        \caption{}
        \label{fig:small1}
    \end{subfigure}
    \hfill
    \begin{subfigure}{0.49\textwidth}
        \centering
        \includegraphics[width=\textwidth]{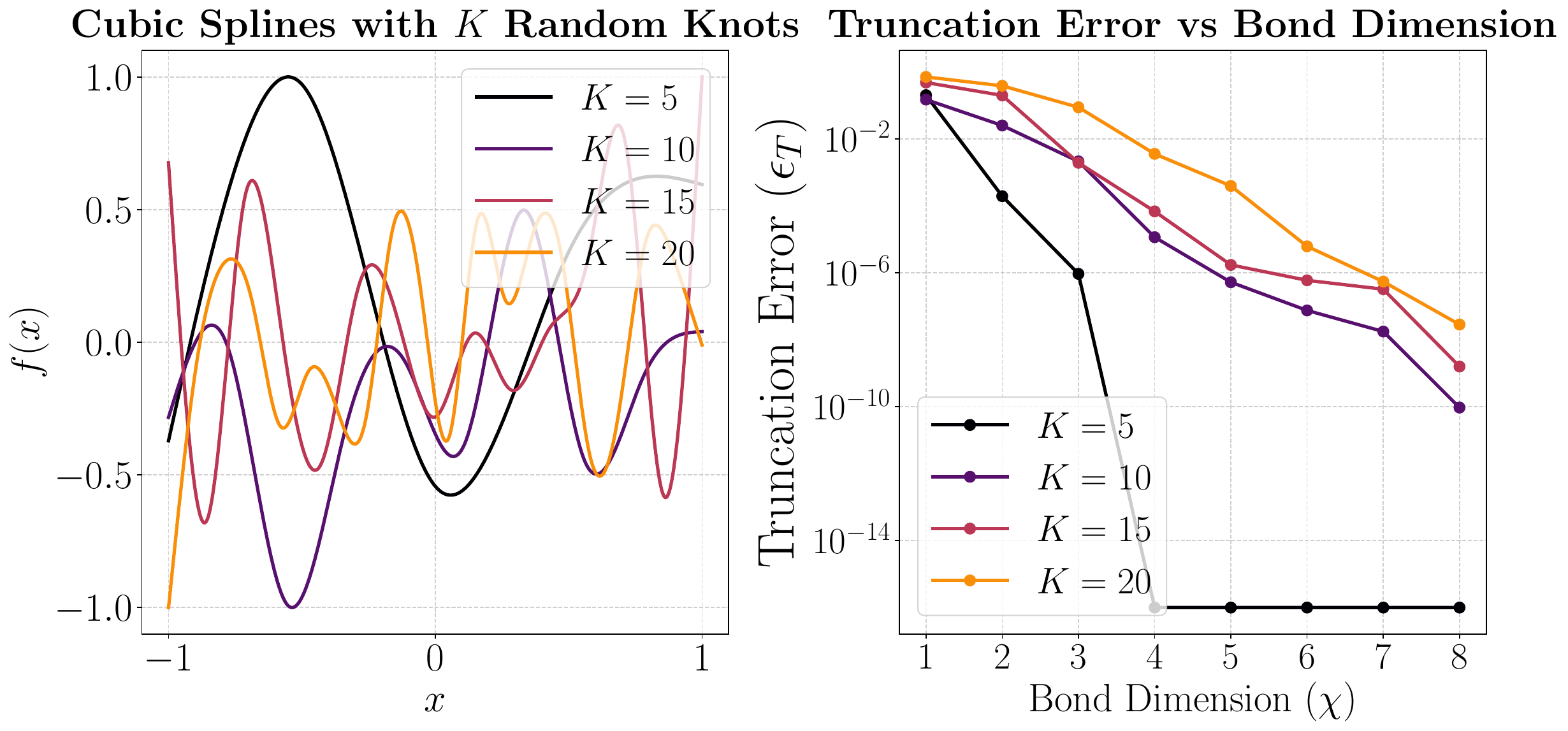}
        \caption{}
        \label{fig:small2}
    \end{subfigure}

    \vspace{0cm} 

    \begin{subfigure}{0.49\textwidth}
        \centering
        \includegraphics[width=\textwidth]{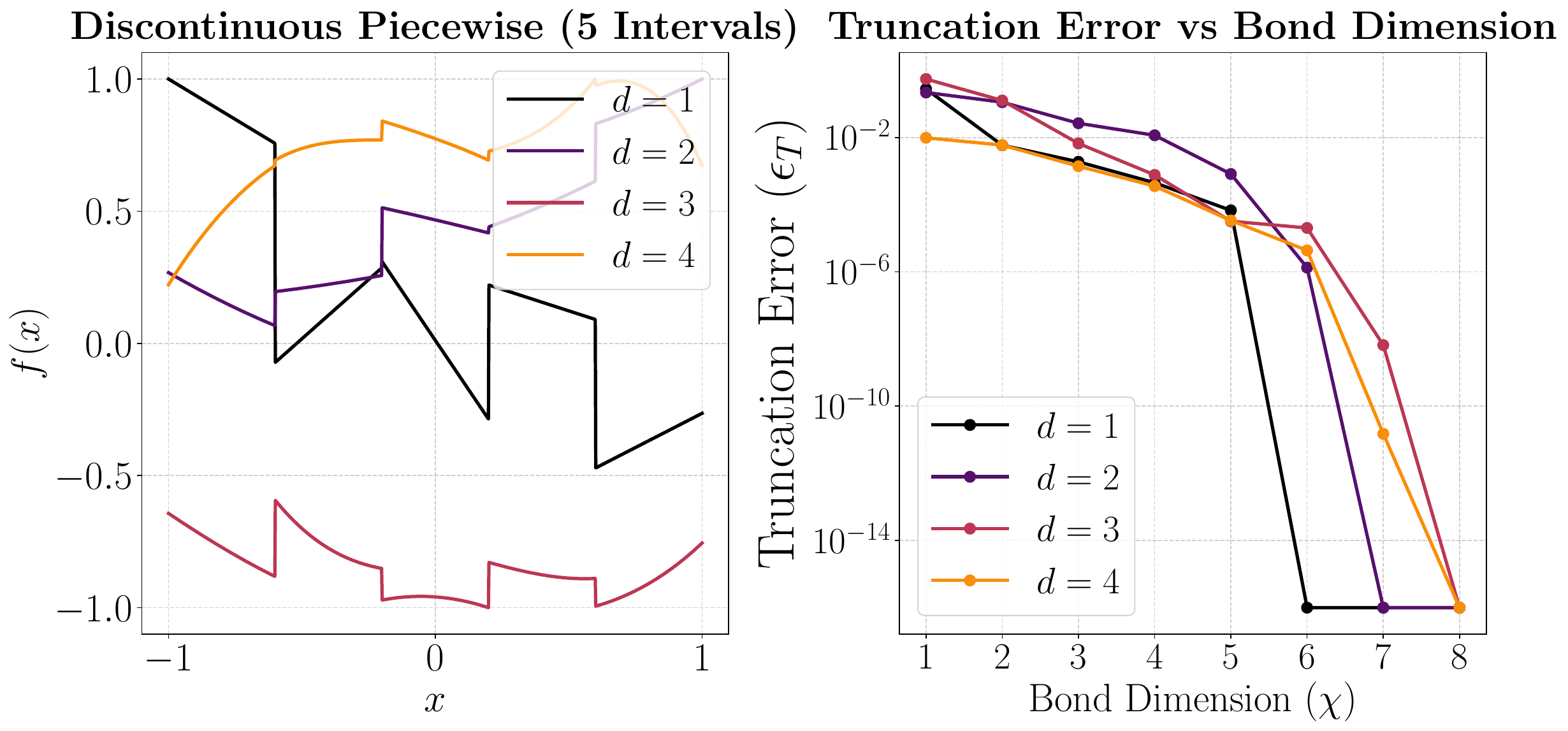}
        \caption{}
        \label{fig:small3}
    \end{subfigure}
    \hfill
    \begin{subfigure}{0.49\textwidth}
        \centering
        \includegraphics[width=\textwidth]{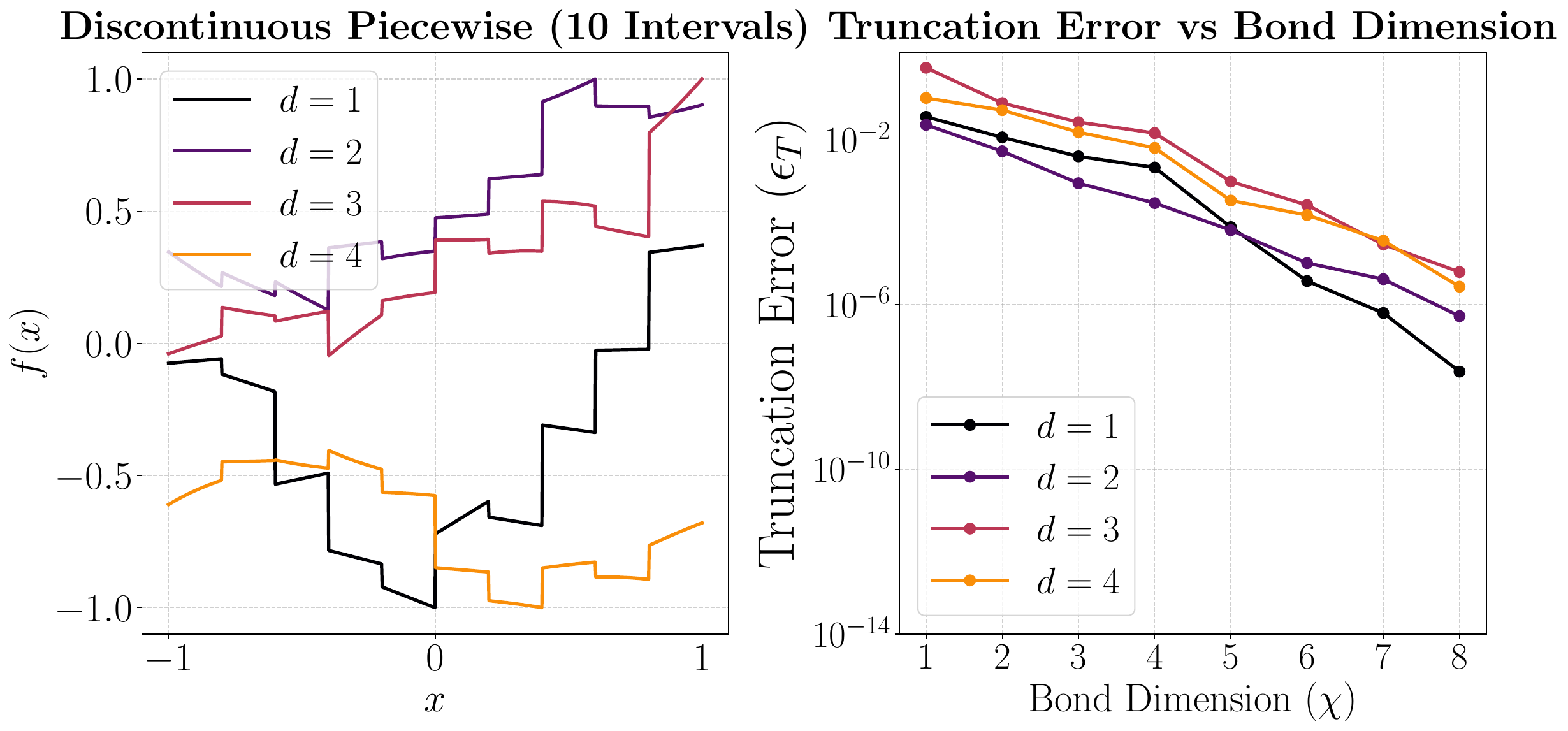}
        \caption{}
        \label{fig:small4}
    \end{subfigure}


    \begin{subfigure}{0.49\textwidth}
        \centering
        \includegraphics[width=\textwidth]{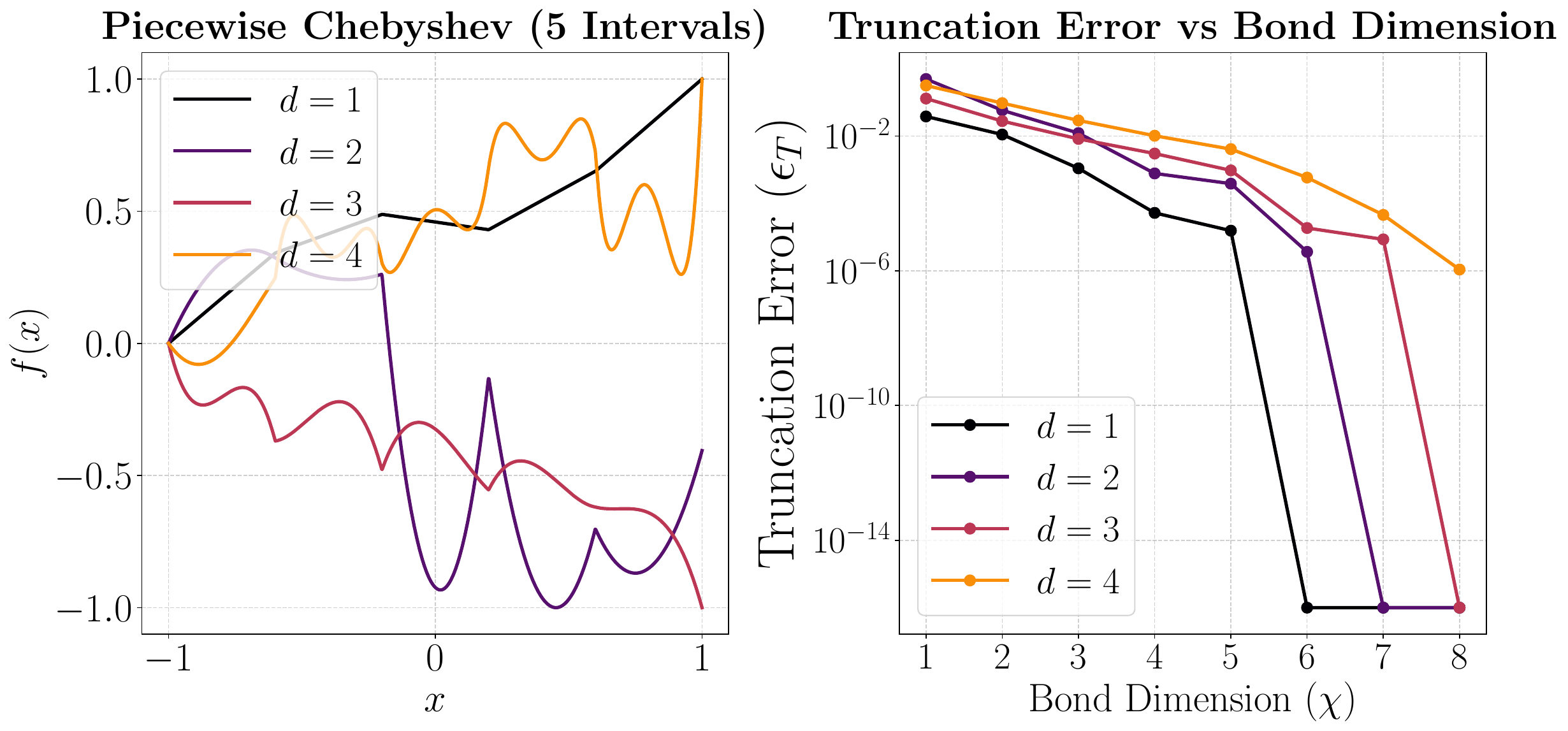}
        \caption{}
        \label{fig:small5}
    \end{subfigure}
    \hfill
    \begin{subfigure}{0.49\textwidth}
        \centering
        \includegraphics[width=\textwidth]{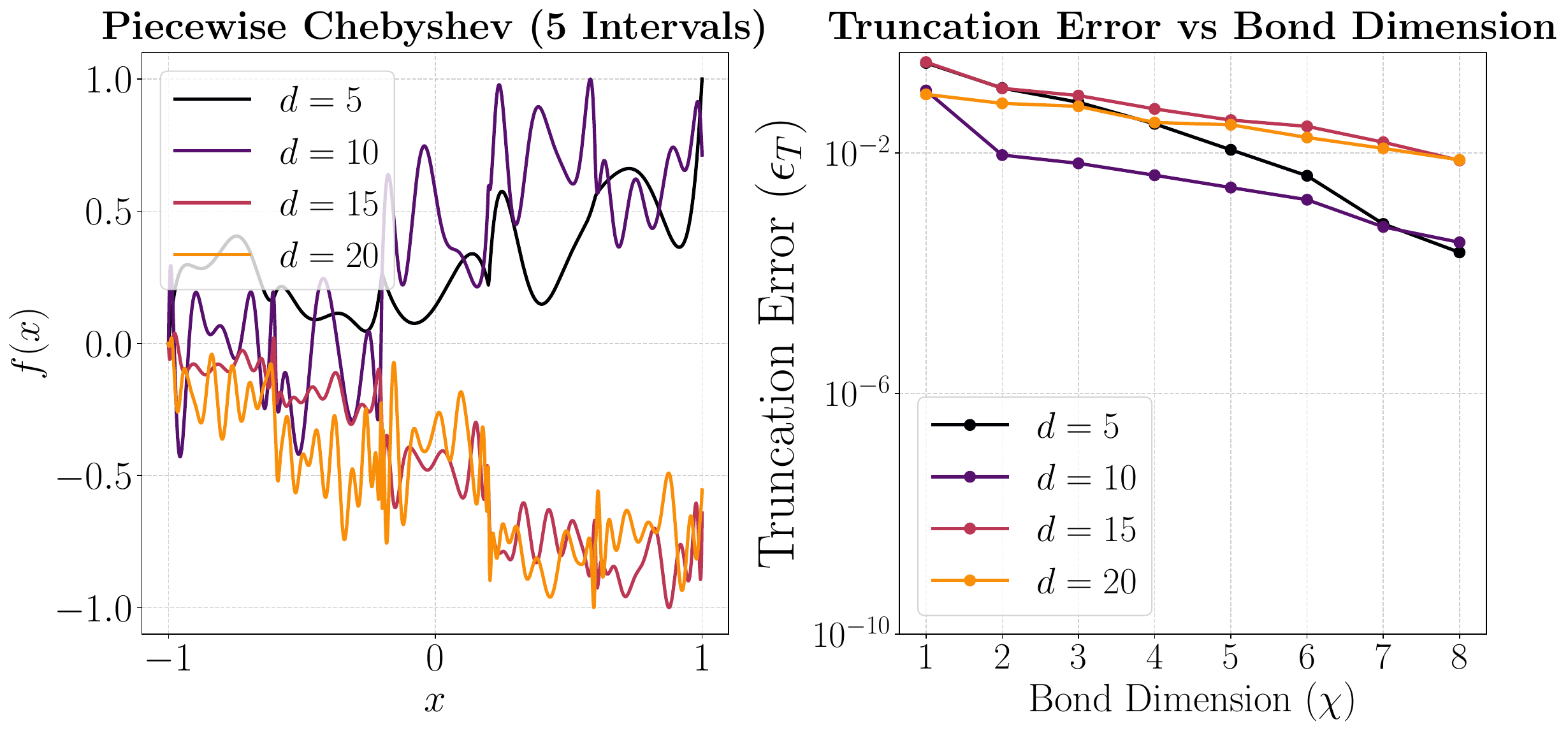}
        \caption{}
        \label{fig:small6}
    \end{subfigure}

    \vspace{0cm} 

    \begin{subfigure}{0.49\textwidth}
        \centering
        \includegraphics[width=\textwidth]{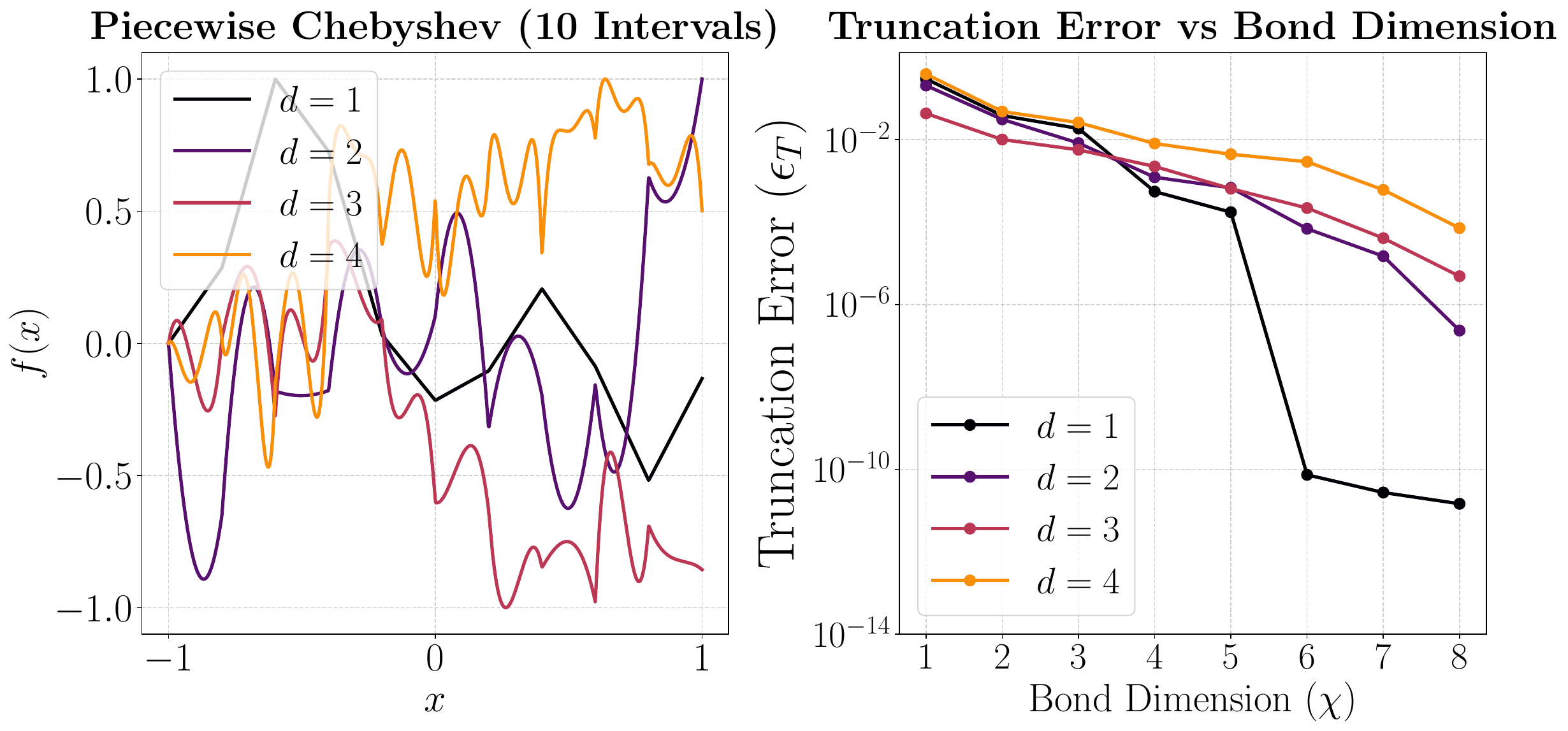}
        \caption{}
        \label{fig:small7}
    \end{subfigure}
    \hfill
    \begin{subfigure}{0.49\textwidth}
        \centering
        \includegraphics[width=\textwidth]{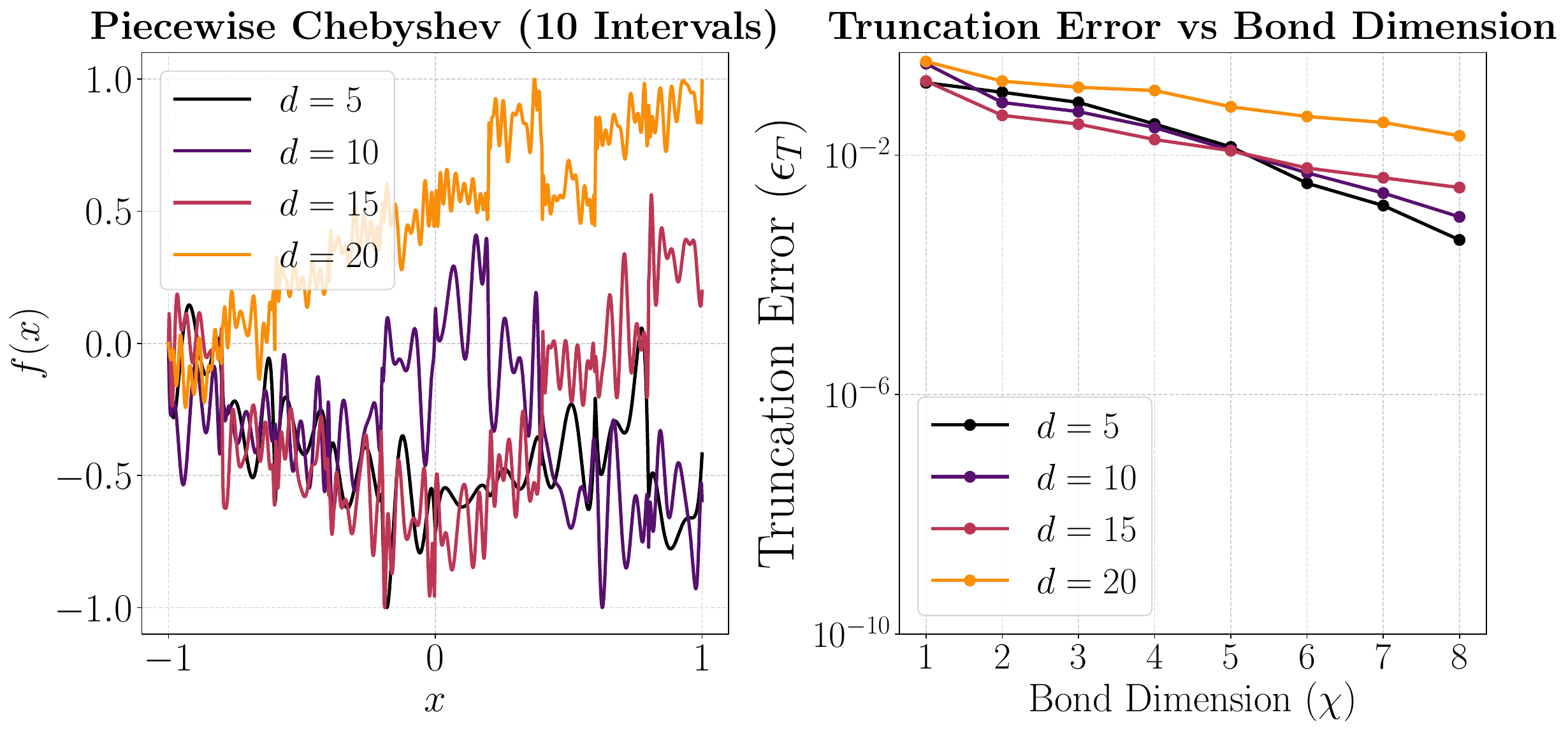}
        \caption{}
        \label{fig:small8}
    \end{subfigure}
    \caption{Truncation Error Analysis for Fixed-$\chi$ Polynomial Functions. 
    The MPS truncation error ($\epsilon_T$) vs $\chi$ ($n=16$) is plotted for (a) Polynomials with uniformly distributed roots, (b) Cubic splines connecting $K$ randomly generated $(x,y)$ points called knots, (c) Discontinuous piecewise polynomials (5 Intervals), (d) Discontinuous piecewise polynomials (10 Intervals), (e) Low-degree piecewise Chebyshev (5 Intervals), (f) High-degree piecewise Chebyshev (5 Intervals), (g) Low-degree piecewise Chebyshev (10 Intervals), (h) High-degree piecewise Chebyshev (10 Intervals).
    All polynomials are generated across $N=65536$ points in the domain [-1,1]. All piecewise polynomials have uniformly distributed piecewise intervals and are generated using randomisation in the coefficients.}
    \label{fig:polynomial_analysis}
\end{figure}

The core results are presented in Figure~\ref{fig:polynomial_analysis} for a class of polynomials of varying functional complexity. The generated functions are included for illustrative purposes, offering an intuitive sense of the general complexity level in each state. This includes standard polynomials, cubic splines, discontinuous piecewise functions, and piecewise Chebyshev polynomials. The polynomial degree $d$ and number of piecewise intervals $I$ are varied widely, neatly capturing the transition between highly structured (low-degree, low-interval) and unstructured (high-degree, high-interval) target vectors. All results were consistent with the theoretical upper bounds derived in \cite{oseledets2013constructive_mps_functions,holmes_smooth_diff_functions}. As anticipated, the truncation error was positively correlated with $d$ and $I$ in nearly all cases. One exception was the class of $I=10$ discontinuous piecewise functions in Figure~\ref{fig:polynomial_analysis}(d), in which the truncation error is essentially independent of $d$. In such states, the degree of discontinuity in the functions is the dominant contribution to entanglement entropy, and the polynomial degree is only a minor contribution. Similar effects were observed in Figure~\ref{fig:polynomial_analysis}(c), though the degree of the polynomial contributes most significantly to the entropy of the state since there are fewer discontinuities in the target vectors. 

The class of moderate to low-degree polynomial functions with limited piecewise intervals can ultimately be approximated with fixed-$\chi$ MPS to a reasonable level of accuracy. This is an encouraging result that MPS is a viable pathway to efficiently representing sufficiently structured information, especially in machine learning contexts where a small degree of noise in the prepared state is permissible and can even offer a natural form of regularisation \cite{vincent2010stacked}. As the information becomes more complex (i.e. described by a high-degree, high-interval Chebyshev polynomial), truncating the bond dimension can induce a consequential degree of error, and an accurate representation necessitates a large bond dimension representation. This presents a critical challenge in the design of efficient quantum circuits to encode these states accurately. 

\subsection*{4.3 Tensor cross interpolation}

Using all $2^n$ coefficients of the state vector to construct the MPS is inefficient for large $n$. Tensor cross interpolation (TCI) enables the approximate construction of MPS states via the strategic sampling of a polynomial-subset of the full state vector \cite{oseledets_tci,PhysRevX.12.041018_tci,PhysRevLett.132.056501_tci,tindall2024compressingmultivariatefunctionstree_tci,SciPostPhys.18.3.104}. We implement the TCI methodology introduced in \cite{oseledets_tci} to assess its efficacy  for the specific function set we encode in the subsequent section. 

We find that the TCI methodology tends to \textit{overestimate} the bond dimension required to construct the MPS representation of a function for a fixed error level. This effect was more pronounced for more complex functions. For example, the construction of the MPS representation of $Cheby(10,10)$ (a degree-10 piecewise Chebyshev polynomial with 10 intervals) required $\chi=16$ to achieve an error of $\epsilon\sim10^{-3}$, whereas a similar error is achievable via the optimal truncation of the exact MPS to just $\chi=8$. Cross approximation constructed this representation using just 80,731 samples, representing just 0.0019\% of the full length of the 32-qubit state vector. This rank overestimation can be accounted for by the greediness of the rank-adaptive loop in the TCI algorithm \cite{oseledets_tci}. 

Rank overestimation in the target state increases the initial computational costs in the MPD algorithm (e.g. by increasing the $\chi_{\text{max}}$ of the initial target MPS, which increases the cost of computing the Schmidt canonical form for compression -- scaling as $\mathcal{O}(n\chi^3)$). However, the overall encoding \textit{performance} of the MPD algorithm is dominated by the quality of the $\chi=2$ approximations at each iteration, which is not affected by the increased rank. The TCI approach is, therefore, naturally compatible with the MPD algorithm and enables the subversion of exponential costs associated with constructing the target MPS from the full state vector. 

\begin{table}[htbp]
    \centering
    \small
    \setlength{\tabcolsep}{6pt}
    \renewcommand{\arraystretch}{1.15}

    \begin{tabular}{l c c c}
        \toprule
        \( f(x) \) & \( \chi \) & \makebox[4cm][c]{Error} & \makebox[3cm][c]{Mean Samples} \\
        \midrule

        \( e^x \) & 1 &
        \makebox[4cm][c]{\( (1.04 \pm 0.01) \times 10^{-15} \)} &
        \makebox[3cm][c]{126} \\

        \( \cos(\pi x) \) & 2 &
        \makebox[4cm][c]{\( (2.72 \pm 2.12) \times 10^{-15} \)} &
        \makebox[3cm][c]{618} \\

        \( x \) & 2 &
        \makebox[4cm][c]{\( (9.23 \pm 4.15) \times 10^{-7} \)} &
        \makebox[3cm][c]{741} \\

        \( x^3 \) & 4 &
        \makebox[4cm][c]{\( (1.05 \pm 0.17) \times 10^{-6} \)} &
        \makebox[3cm][c]{8065} \\

        \( N(0,1^2) \) & 4 &
        \makebox[4cm][c]{\( (8.03 \pm 1.03) \times 10^{-7} \)} &
        \makebox[3cm][c]{1994} \\

        \( N(0,0.3^2) \) & 7 &
        \makebox[4cm][c]{\( (2.09 \pm 0.48) \times 10^{-6} \)} &
        \makebox[3cm][c]{8739} \\

        \(\text{Cheby}(10,1)\) & 7 &
        \makebox[4cm][c]{\( (8.81 \pm 0.06) \times 10^{-7} \)} &
        \makebox[3cm][c]{15485} \\

        \(\text{Cheby}(10,4)\) & 13 &
        \makebox[4cm][c]{\( (1.71 \pm 0.18) \times 10^{-6} \)} &
        \makebox[3cm][c]{55394} \\

        \(\text{Cheby}(10,10)\) & 16 &
        \makebox[4cm][c]{\( (9.96 \pm 0.34) \times 10^{-4} \)} &
        \makebox[3cm][c]{80731} \\

        \bottomrule
    \end{tabular}

    \caption{\textbf{Cross-Approximation Error Metrics (}$\mathbf{n=32}$\textbf{ Qubits).} \newline
    All functions are defined over the domain $D=[-1,1]$. $\text{Cheby}(I,d)$ represents a degree-$d$ piecewise Chebyshev polynomial with $I$ uniform piecewise intervals. The ideal maximum bond dimension was determined based on preliminary testing and 4 repeated trials were computed for each function. Uncertainty in error represents standard deviation.}
    \label{tab:cross_approximation_data}
\end{table}

\subsection*{4.4 Quantum Function Encoding Results}

As a central point of comparison, the TNO strategy based on the scheme introduced by \cite{melnikov} was tested by encoding the sine function on 12 qubits across 5 layers of the hardware efficient ansatz (HEA) with up to 500 iterations. The sine function was found to be encoded with fidelity $\sim 10^{-4}$. While this is still a high-quality state, the sine function has an exact $\chi=2$ representation for any number of qubits \cite{oseledets2013constructive_mps_functions}. This means that the MPD algorithm~\cite{ran} can prepare this state to machine precision with a single layer of arbitrary $U(4)$ gates with an extremely minimal computational cost. Crucially, the HEA-based TNO uses a random initialisation of parameters, whereas the sequential MPD+TNO explored in this paper has the advantage of using the MPD algorithm to initialise parameters. We find that the random parameter initialisation significantly increases the likelihood of low-quality local minima in the training of the HEA-based tensor network ansatz when compared to the relatively high-quality initialisation of parameters via the MPD algorithm. This result was established for a point of comparison to the MPD+TNO algorithm to illustrate the significant role of parameter initialisation in determining training efficiency. 


\begin{figure}[htbp]
    \centering

    \begin{subfigure}{0.32\textwidth}
        \centering
        \includegraphics[width=\textwidth]{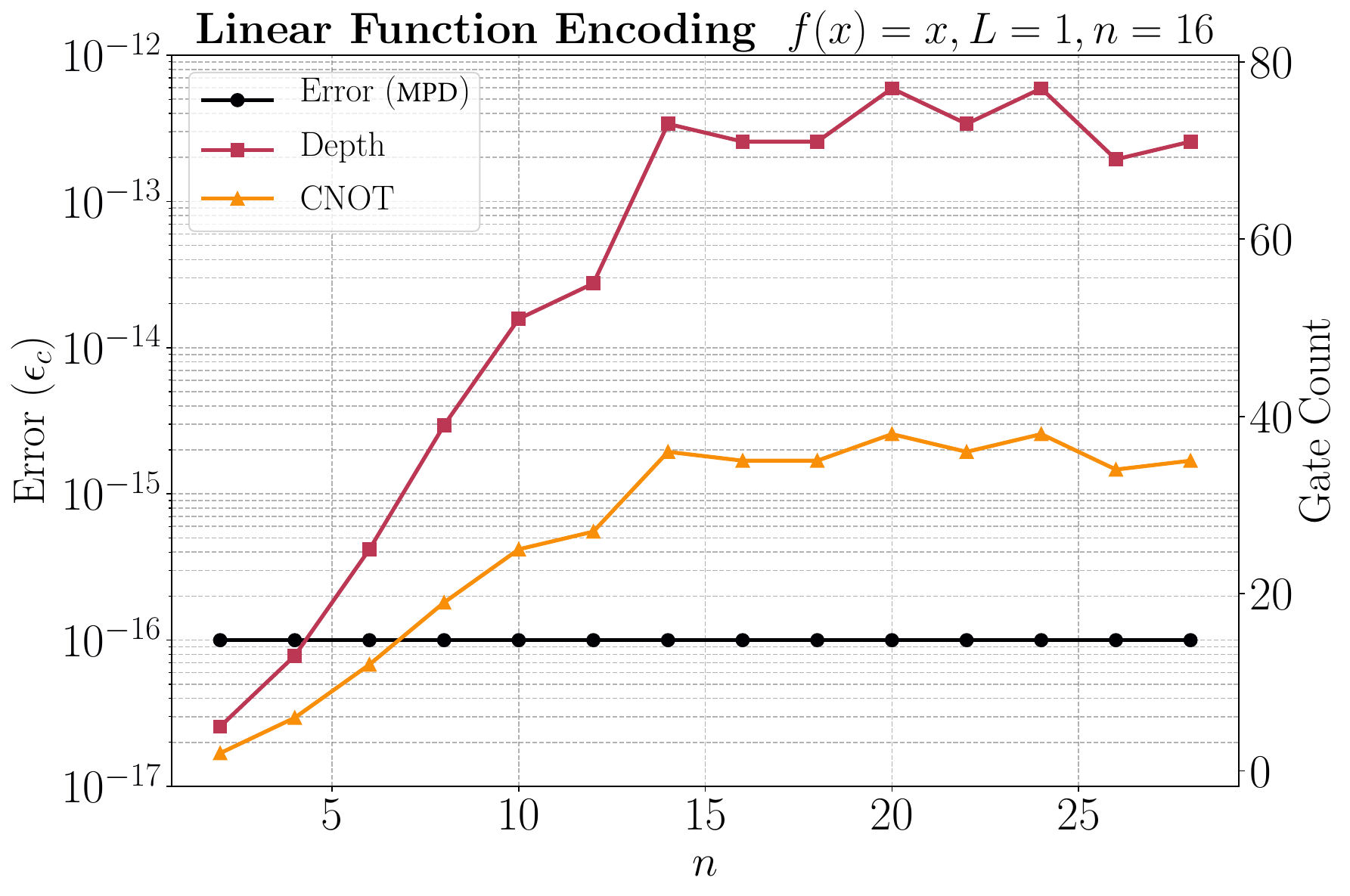}
        \caption{}
        \label{fig:small1}
    \end{subfigure}
    \hfill
    \begin{subfigure}{0.32\textwidth}
        \centering
        \includegraphics[width=\textwidth]{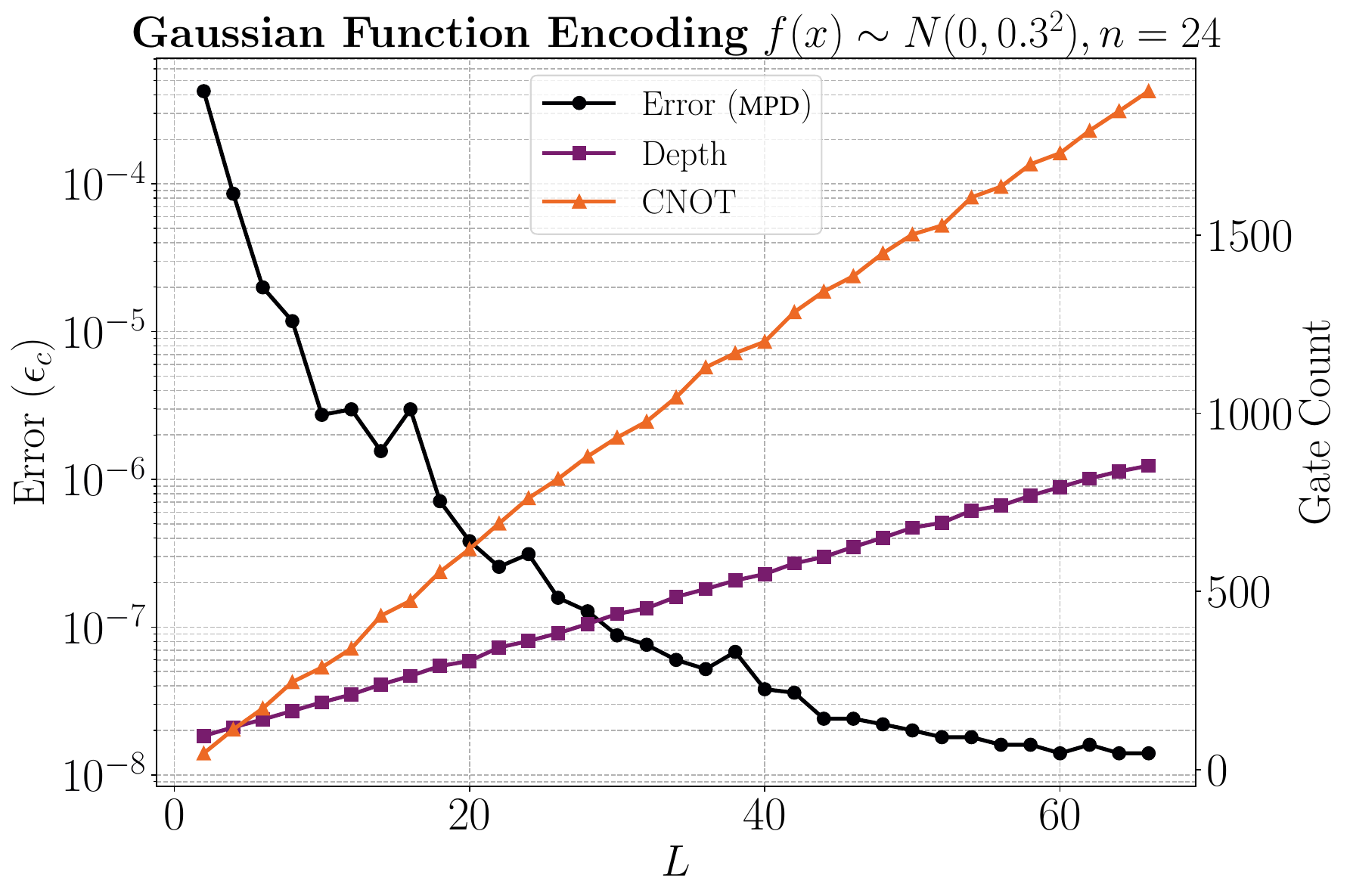}
        \caption{}
        \label{fig:small2}
    \end{subfigure}
    \hfill
    \begin{subfigure}{0.32\textwidth}
        \centering
        \includegraphics[width=\textwidth]{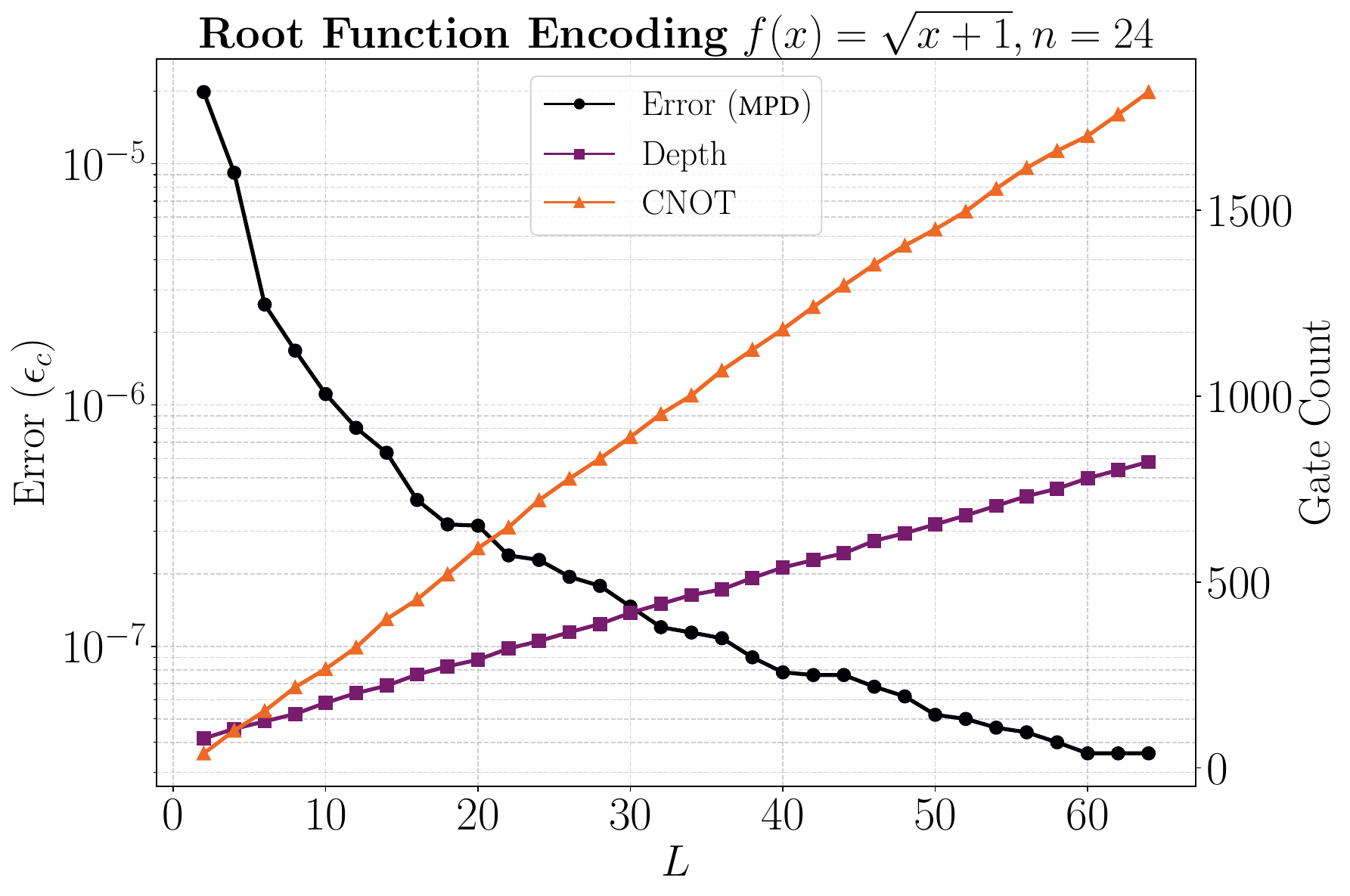}
        \caption{}
        \label{fig:small3}
    \end{subfigure}

    \vspace{0.1cm} 

    \begin{subfigure}{0.32\textwidth}
        \centering
        \includegraphics[width=\textwidth]{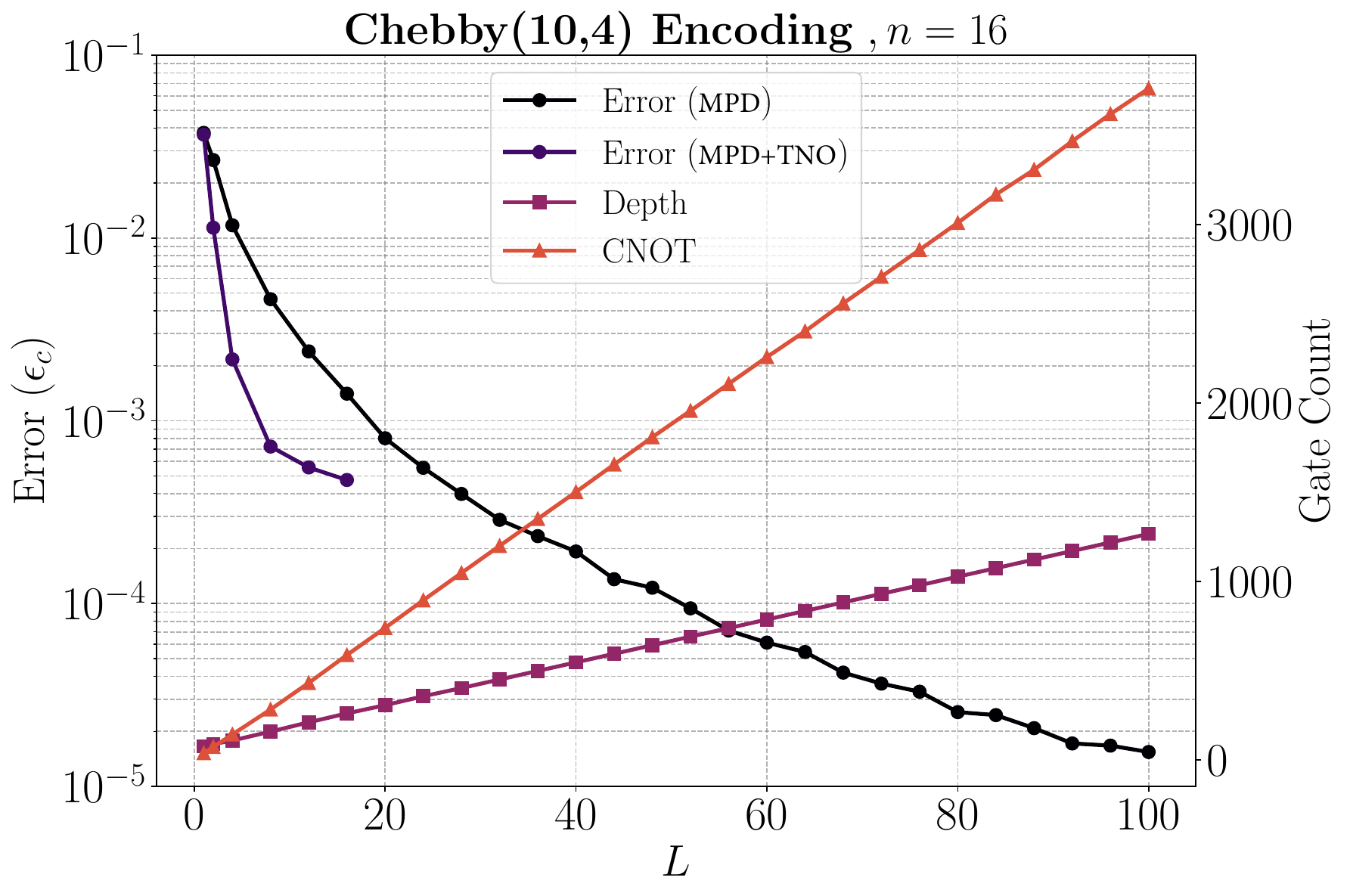}
        \caption{}
        \label{fig:small4}
    \end{subfigure}
    \hfill
    \begin{subfigure}{0.32\textwidth}
        \centering
        \includegraphics[width=\textwidth]{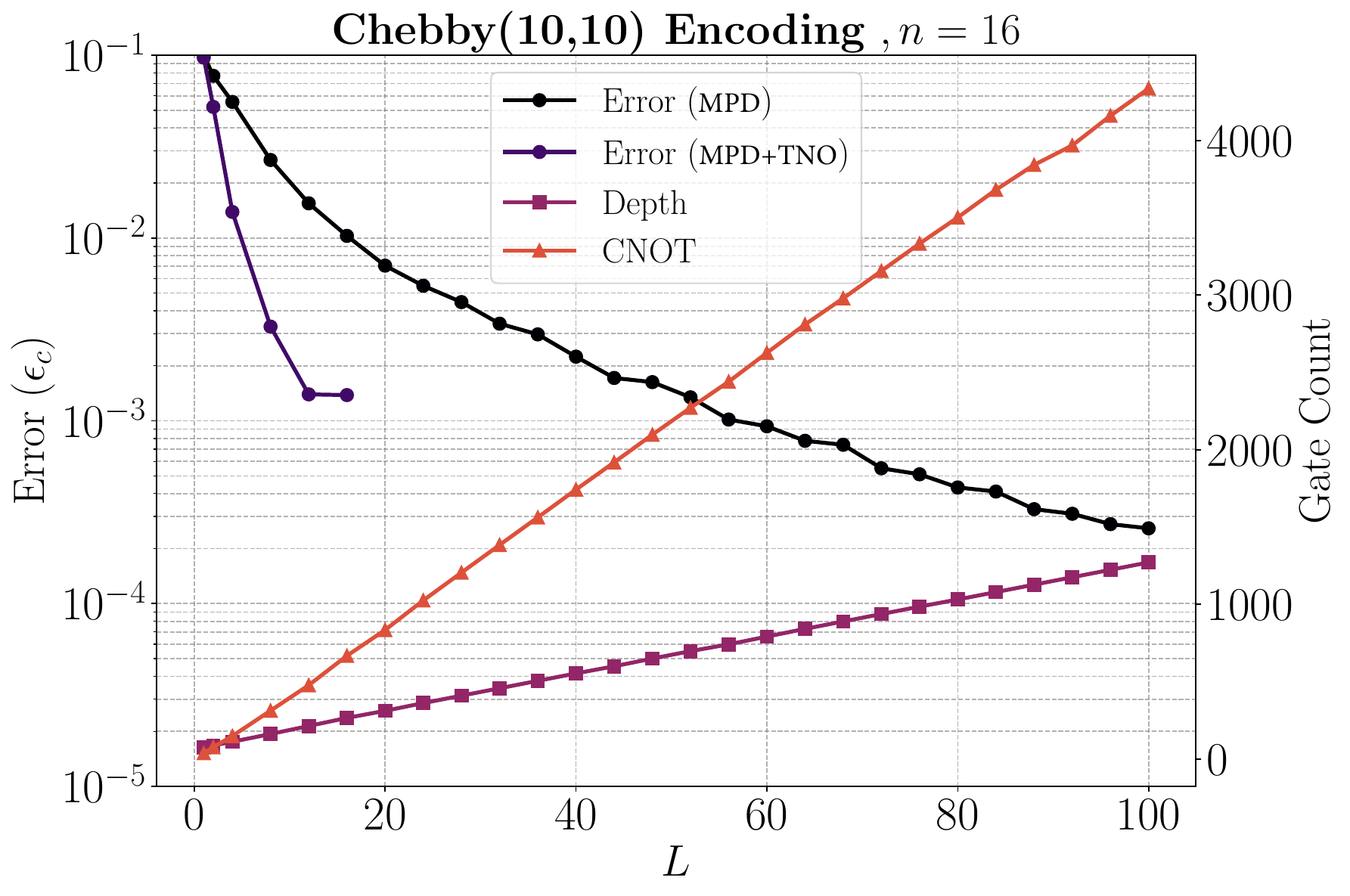}
        \caption{}
        \label{fig:small5}
    \end{subfigure}
    \hfill
    \begin{subfigure}{0.32\textwidth}
        \centering
        \includegraphics[width=\textwidth]{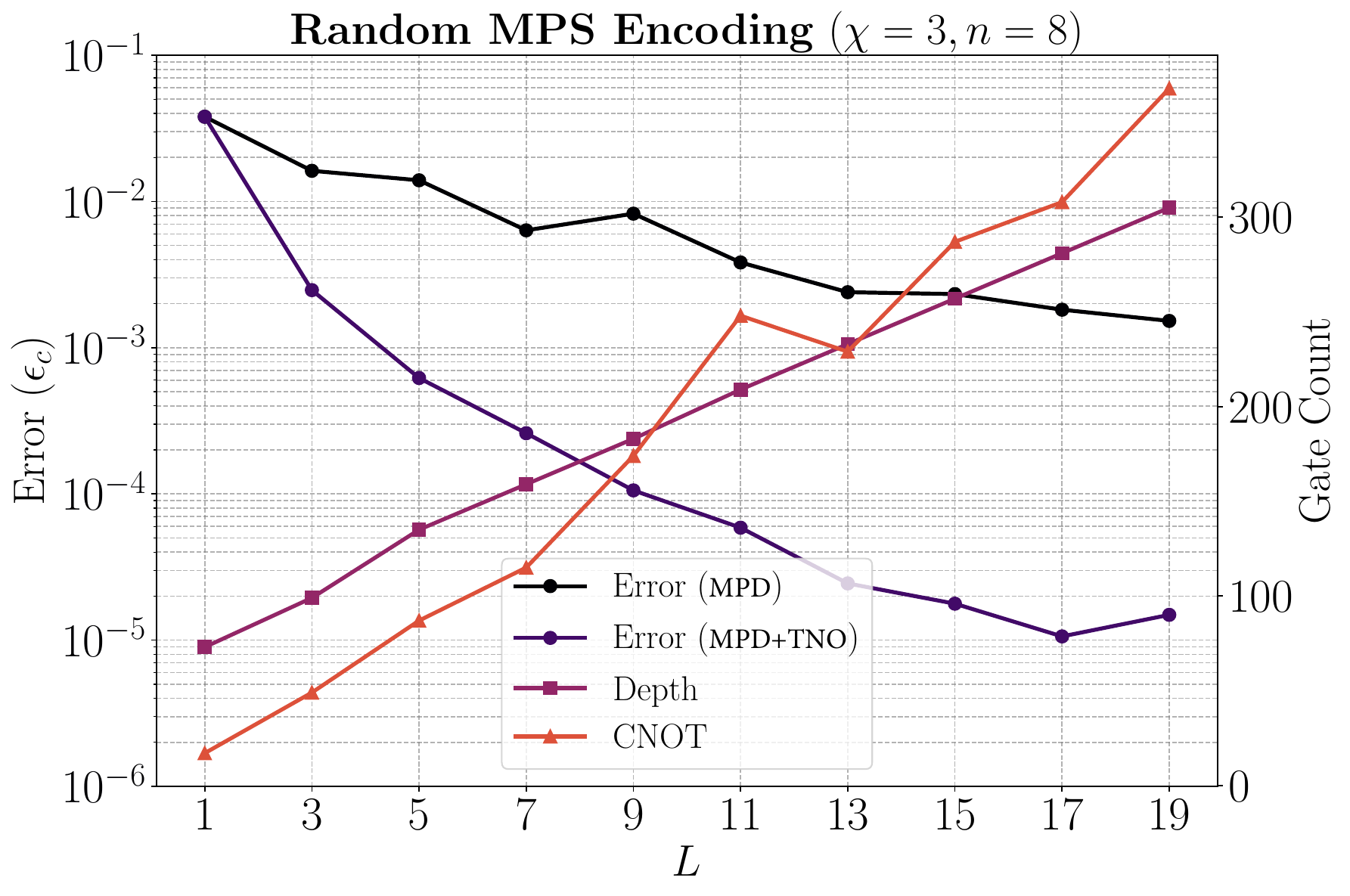}
        \caption{}
        \label{fig:small6}
    \end{subfigure}

    \vspace{0.1cm} 

    \begin{subfigure}{0.32\textwidth}
        \centering
        \includegraphics[width=\textwidth]{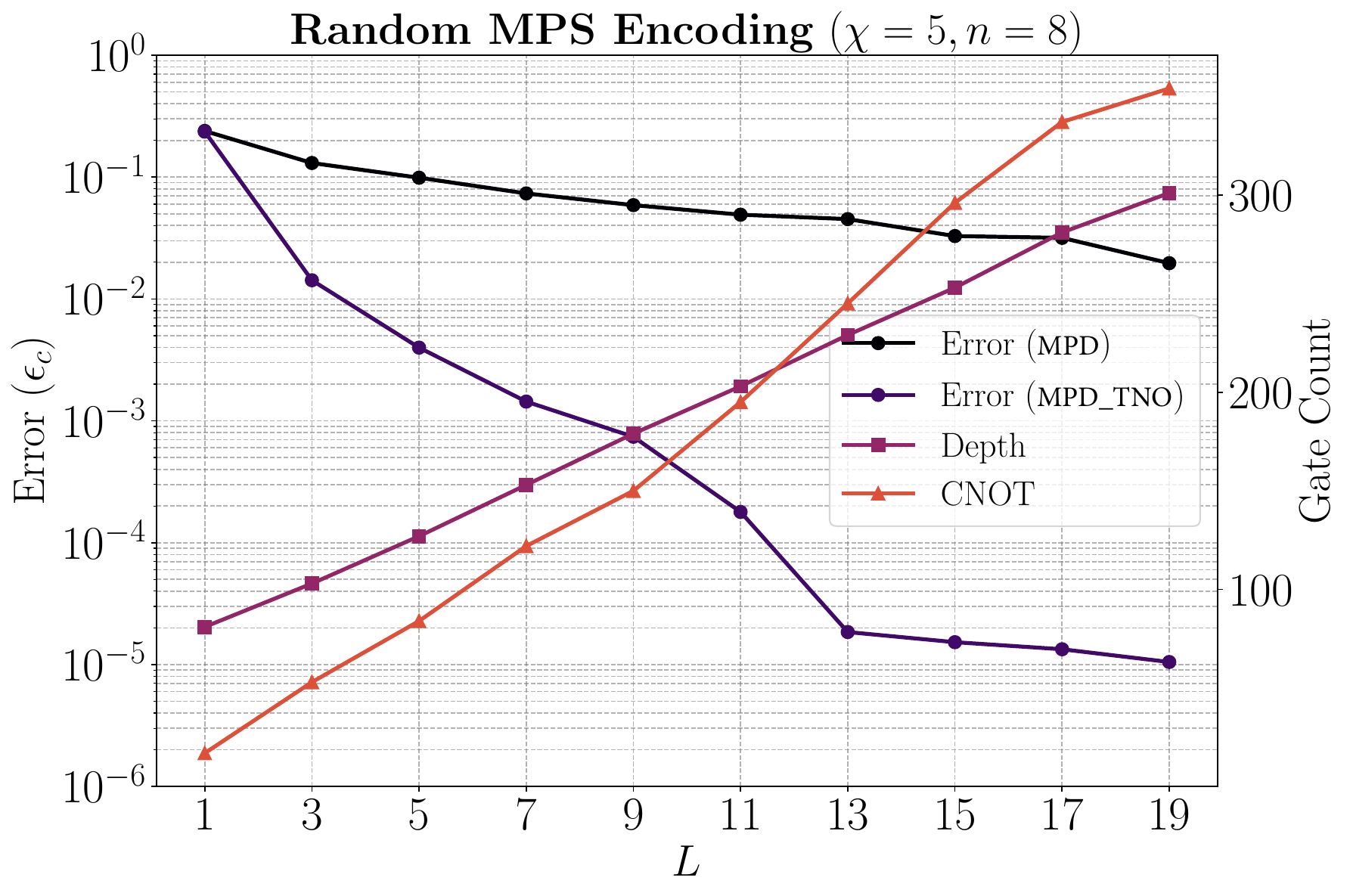}
        \caption{}
        \label{fig:small7}
    \end{subfigure}
    \hfill
    \begin{subfigure}{0.32\textwidth}
        \centering
        \includegraphics[width=\textwidth]{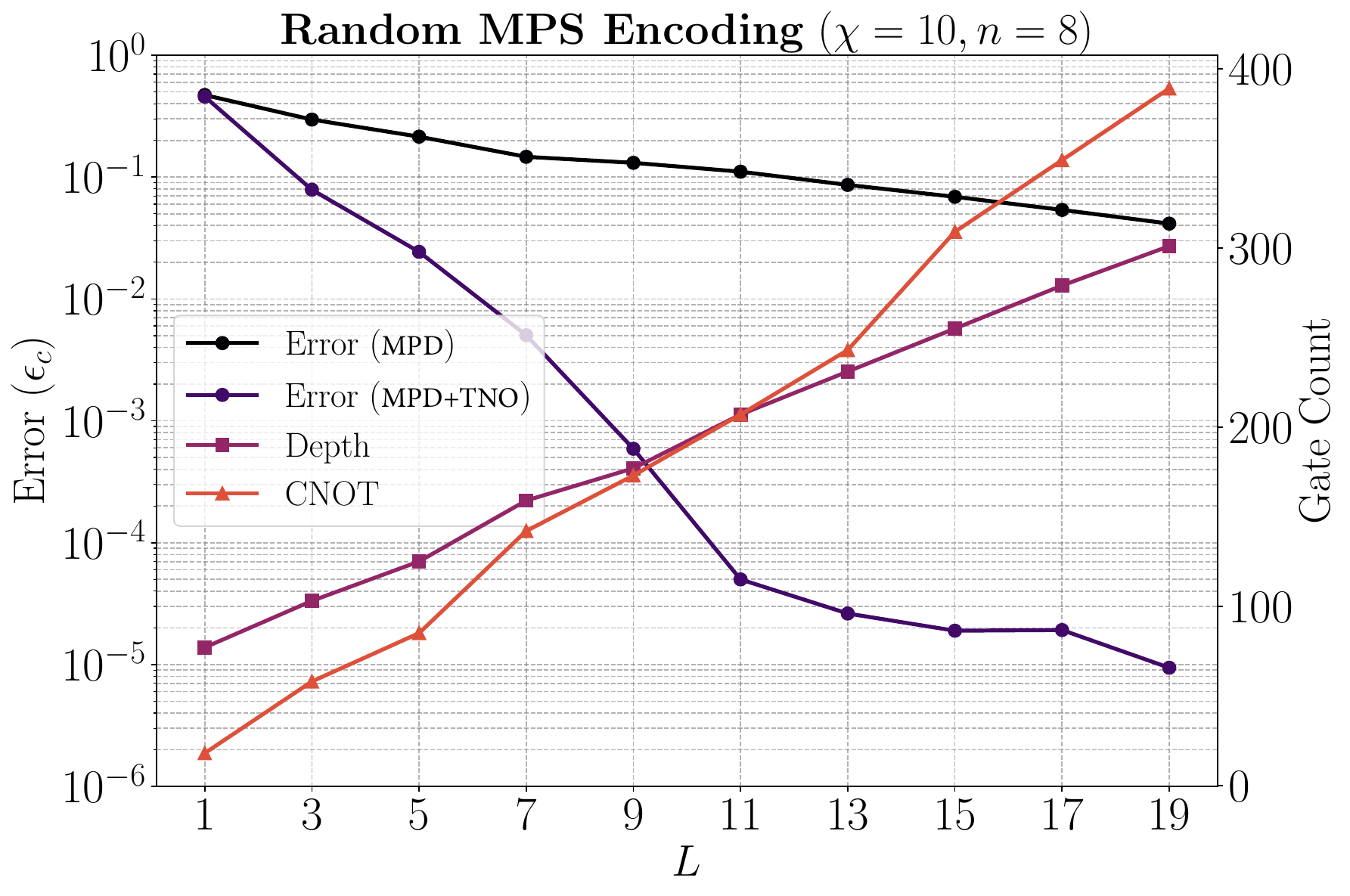}
        \caption{}
        \label{fig:small8}
    \end{subfigure}
    \hfill
    \begin{subfigure}{0.32\textwidth}
        \centering
        \includegraphics[width=\textwidth]{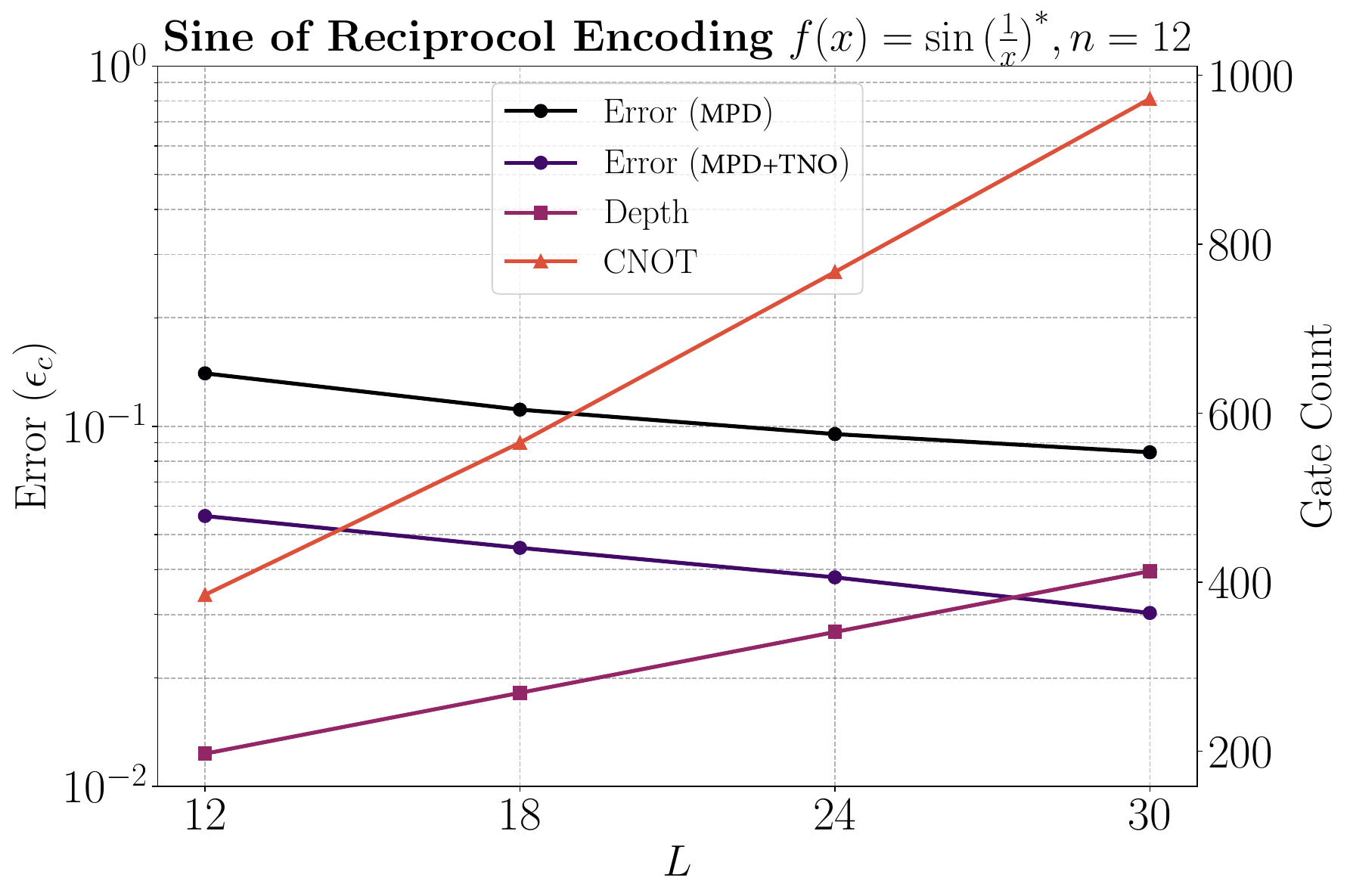}  
        \caption{}
        \label{fig:small9}
    \end{subfigure}

    \caption{Benchmarks of the MPD and MPD+TNO Algorithms. $Cheby(I,d)$ denotes a randomly generated degree-$d$ Chebyshev polynomial with $I$ piecewise intervals. All functions were generated in the domain $D=[-1,1]$. The CNOT and circuit depth gate counts were retrieved using the Qiskit decomposition of the arbitrary $U(4)$ unitaries into elementary quantum gates. 500 iterations were used for the MPD+TNO algorithm.}
    \label{fig:encoding_analysis}
\end{figure}

The quantum circuits produced by the MPD and MPD+TNO algorithms were simulated for a range of target states, and the results are presented in Figure~\ref{fig:encoding_analysis}. In all trials, we use an adaptive bond dimension by truncating all singular values below a threshold of $\epsilon=1\text{e}^{-10}$. The adaptive bond dimension helps to balance the computational costs with the accuracy of the prepared state. Alternatively, one could set a fixed $\chi_{\text{max}}$ during the computational and training process. We also note that we construct the exact MPS before truncation. For large numbers of qubits, we could construct the MPS via efficient cross-approximation, which has time complexity $\mathcal{O}(n\chi_{\text{max}}^2)$ \cite{oseledets2010fast_TT_cross}, without the need for the full state vector to be specified. We summarise the results of the discretised function encoding circuits and the performance of the MPD and MPD+TNO algorithms with four key points:

\textbf{1.}  The MPD scheme provides efficient, low-depth, and accurate circuits that prepare discretised functions with accurate low-bond dimension representations. This included a single-layer exact encoding of the linear line function in Figure~\ref{fig:encoding_analysis}(a), the Gaussian $f(x)\sim N(0,0.3^2)$ in Figure~\ref{fig:encoding_analysis}(b), the root function $f(x)=\sqrt{x+1}$ in Figure~\ref{fig:encoding_analysis}(c), and low-degree piecewise polynomials in Figure~\ref{fig:encoding_analysis}(d)-(e). The fidelity of the encoded target state can be tuned by adjusting the number of layers $L$. However, the fidelity that can be achieved inevitably degrades as a larger bond dimension is required to represent the state accurately. Nonetheless, the MPD encoding scheme was numerically stable and efficient in classically generating quantum circuits up to a modest number of layers. The generated circuits are NISQ-friendly with CNOT gate complexity and circuit depth shown to scale linearly with $L$, aligning with the expected $\mathcal{O}(nL)$ circuit depth scaling.

\textbf{2.}  The MPD algorithm generalises poorly beyond the context of the state preparation of low-bond dimension functions. This is because the MPD algorithm's accuracy and rate of convergence depend on the fidelity of the $\chi=2$ approximation at each iteration. For example, the state vector discretising $f(x)=\sqrt{x+1}$ has a $\chi=2$ approximation with fidelity of $F=0.99998\pm0.00001$. On the other hand, truncating a random $\chi=5$ MPS to $\chi=2$ has a fidelity of $F=0.36\pm 0.11$ (10 trials). This is because random MPS tend to have fairly uniform singular values compared to the exponentially decaying singular values for well-behaved smooth functions. The performance of the MPD algorithm was inadequate in preparing sufficiently high-quality encodings of all random MPS explored, as well as the irregularly behaving $f(x)=\sin{(\frac{1}{x})}$ function in Figure~\ref{fig:encoding_analysis}(i). In these states, the fidelity improves only marginally with increasing layers. This is consistent with earlier observations made by \cite{BenDov2024,rudolph2022decompositionmatrixproductstates} that the analytic MPD algorithm exhibits diminishing marginal gains in fidelity at each subsequent layer. 
It is important to note that the classical computational cost grows exponentially with \( L \), making it computationally intractable to compute arbitrarily deep circuits to capitalise on these marginal gains. 

\textbf{3.}  The MPD+TNO scheme significantly improves the fidelity of the prepared target states. This effect is especially pronounced when the analytic MPD scheme performs poorly, as illustrated in the random MPS in Figure~\ref{fig:encoding_analysis}(e)-(h). It was also shown to improve the fidelity of the piecewise Chebyshev functions Figure~\ref{fig:encoding_analysis}(d)-(e) at low circuit depths (this was not extended to larger numbers of layers due to the high qubit count contributing to significant computational costs). This effect can be explained by the observation that that expressible $\chi$ grows as $\chi=\mathcal{O}(2^\chi)$ \cite{ran} with the exact encoding of a fixed-$\chi$ MPS requiring $\mathcal{O}(n\chi^2)$ circuit depth in general \cite{schon_sequential_MPS_generation}. 
In many instances, deeper quantum circuits from the MPD algorithm lead to significantly higher fidelity approximations of the target state.
The numerical results support this finding, demonstrating that the MPD+TNO scheme can significantly improve performance. This enables an improvement in the fidelity of the target state without increasing the circuit depth, albeit at the expense of an additional classical cost in optimising the tensor network representation of the quantum circuit.
    
\textbf{4.}  The MPD+TNO scheme suffers from trainability problems as the number of qubits increases. The optimisation is generally efficient when the initialised parameters overlap significantly with the global minimum. However, the algorithm's overall performance degrades as the dimension of the parameter landscape increases, corresponding to deeper circuits with a greater number of qubits. Alternative approaches to gradient descent and optimisation strategy could potentially be considered to improve trainability for deeper circuits. For instance, it was shown by Rudolph et al. \cite{rudolph2022decompositionmatrixproductstates} that iteratively optimising layer-by-layer (rather than all the layers at once) can improve training efficiency in some instances. This alternative approach comes with a greater (quadratic) dependence of complexity on the number of iterations scaling as $\mathcal{O}(nT^2L\chi_{\text{max}}^3)$ \cite{rudolph2022decompositionmatrixproductstates}. Alternatively, training efficiency could be significantly reduced by truncating $\chi_{\text{max}}$ to a reasonable target level, especially when $n$ is large. 

\section*{5. MPS Preparation of Images}

\subsection*{5.1 Motivation}

Although there is no definitive evidence of a quantum advantage in this domain, there has been extensive exploration into the potential of quantum machine learning for structured classical data \cite{Cerezo_challenges_and_opportunities_in_QML,Rath2024,Senokosov_2024_image_class,Landman_2022_qml_image,Guala2023_qml_image,compressed_prep_for_QML,PhysRevA.111.012630_qml_img_tn}. This includes quantum models for the classification of images \cite{Senokosov_2024_image_class,Landman_2022_qml_image,Guala2023_qml_image,compressed_prep_for_QML,PhysRevA.111.012630_qml_img_tn}. Beyond the context of image encoding in the training of quantum classification models, quantum image encoding has also been proposed as a subroutine in quantum edge detection \cite{Yuan2019_qntm_edge_Detection,Llorens_2025_qntnm_edge_detection}.

The most straightforward approach to quantum image encoding is the mapping of pixel values into the angles of rotation gates \cite{Rath2024}. This mapping provides both a simple non-linear feature map and the embedding of data into the exponential-dimensional Hilbert space, which are attractive properties for quantum kernel methods. However, angle-encoding requires the number of qubits in the system to scale linearly with the number of pixels in the image, which prohibits scaling this approach to even modestly-sized images. Alternatively, the Flexible Representation of Quantum Images (FRQI) \cite{Le2011_frqi} and the Novel Enhanced Quantum Representation of Digital Images (NEQR) \cite{Zhang2013_neqr} provide quantum image encoding protocols for amplitude and basis-encoding approaches, respectively. However, these methods require a gate count that scales with the number of pixels in the image. MPS-based encoding represents a potential alternative to these conventional approaches, incorporating image compression through circuits of depth scaling at most $\mathcal{O}(n\chi^2)$ where the MPS bond dimension $\chi$ serves as a compression hyperparameter. The approximate MPS state is still encoded into a Hilbert space of dimension at least equal to the number of pixels in the image.

While being outperformed by more expressive tensor networks like tree tensor networks (TTNs) and the multi-scale renormalisation ansatz (MERA) tensor networks, classical MPS-based tensor network classifiers are sufficiently expressive to achieve a $>$99.0\% accuracy in classifying the MNIST data set \cite{NIPS2016_5314b967_ctnml}. However, more complex image datasets obey or may even exceed two-dimensional area laws of entanglement entropy scaling, which is outside the theoretically expressible scope of one-dimensional MPS with bounded bond dimensions \cite{lu_tn_efficient_descriptions_classical_data}. Recent research has also explored using tensor network optimisation of the sequential structure depicted in Figure~\ref{fig:ran_circuit} (without the MPD initialisation) for the amplitude encoding of the Fourier coefficients of image data \cite{Jobst2024efficientmps}. The authors in \cite{Jobst2024efficientmps} compare the sequential MPS ansatz to a MERA and a two-dimensional sequential ansatz, finding that all circuits exhibit similar performance. While this is a strong indication of the expressive power of the one-dimensional sequential ansatz, it may be that MPS are well-suited to capturing the relatively straightforward entanglement structure of the $28\times 28$ Fashion-MNIST dataset used in the study, rather than a clear indication of the viability of MPS in capturing more general and complex datasets. 

\subsection*{5.2 Approximating Images with MPS}

\begin{figure}[htb]
    \centering
    \begin{subfigure}[b]{\textwidth}
        \centering
        \includegraphics[trim=0cm 0cm 0cm 0cm, clip, width=0.95\textwidth]{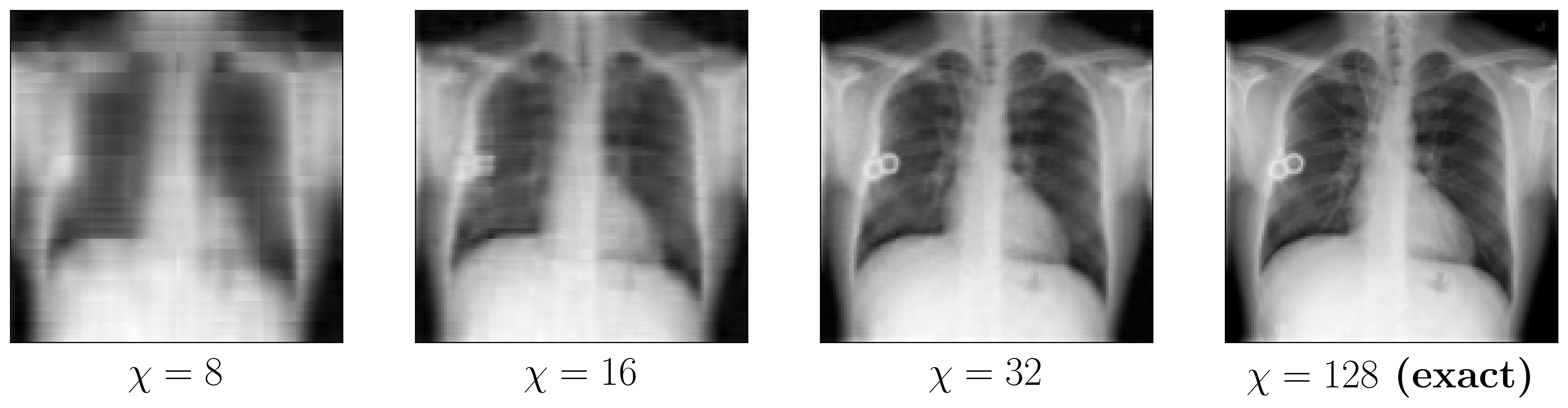}
        \caption{} 
        \label{fig:subfig_a}
    \end{subfigure}
    
    \vspace{1em} 

    \begin{subfigure}[b]{\textwidth}
        \centering
        \includegraphics[trim=0cm 0cm 0cm 0cm, clip, width=0.95\textwidth]{New_Image_Encoding_Results/visualising_image_singular_values_updated_shortened.png}
        \caption{} 
        \label{fig:subfig_b}
    \end{subfigure}
    
    \caption{(a) Image compression through bond dimension truncation for a $128\times 128$ ChestMNIST image. (b) Visualising the singular value (Schmidt coefficient) spectrum across the central virtual indices $\alpha_4$ through $\alpha_{10}$ in the exact $n=14$ representation of the above target image. The central virtual index $\alpha_7$ is of dimension $\alpha_7=128$. Truncation of the exact $\chi=128$ bond dimension to $\chi'$ corresponds to retaining at most $\chi'$ singular values in the Schmidt decomposition across each virtual index, with the error corresponding to the root of the square of discarded singular values. The rapid decay of singular values bounds the error when approximating the target image by a smaller bond dimension.}
    
    \label{fig:mps_compression_graphic}
\end{figure}

We adopt the $128\times 128$ ChestMNIST dataset as a case study. This dataset moderately extends beyond the widely used MNIST and Fashion-MNIST in both complexity and size. Approximating images using MPS with reduced bond dimension is comparable to a direct rank-reduced image approximation. We show the image quality for varying reduced-$\chi$ MPS approximations for the first image in the ChestMNIST dataset in Figure~\ref{fig:mps_compression_graphic}(a), illustrating that the approximation quality and degree of compression can be tuned via bond dimension truncations. We can also see in Figure~\ref{fig:mps_compression_graphic}(b) that the singular values corresponding to each of the central virtual indices decay rapidly, providing an intuition behind why the $\chi$-reduced truncation error, defined in terms of the Frobenius norm error in Equation~\ref{eq:truncation_error}, remains small. Additionally, the $\chi=2$ approximation of this target image yields a fidelity of $F=0.8971$, which is the exact fidelity achievable with a single-layer of $U(4)$ gates.

\subsection*{5.3 Image Encoding Methodology}

We consider the representation of a target image $T=T(i,j)$ of size $2^{\frac{n}{2}}\times 2^{\frac{n}{2}}$ using MPS. We vectorise the image by flattening the normalised pixel values into a vector of dimension $2^{n}$ using row-by-row indexing.  The quantum state represented by this target state vector corresponds to:
\begin{equation}
    \ket{T} = \sum_{i,j=0}^{2^{\frac{n}{2}-1}} T_{ij} \ket{i}\ket{j}
\end{equation}
where $T_{ij}$ corresponds to the normalised pixel value of the target image $T$ at position $(i,j)$. This is a convenient format since the basis states correctly index the pixel values.

We construct an MPS with $n$ sites using the exact construction algorithm \cite{gundlapalli_deterministic_MPS_preparation} provided in Algorithm 1. This process enables the bond dimension of the target image to be truncated to a desired approximation quality. The approximation quality for a given bond dimension is problem-dependent. However, we find that the MPS format is surprisingly well-suited to representing the ChestMNIST dataset. For example, the truncation of the exact representations of the ChestMNIST dataset images from the original bond dimension of $\chi=128$ to $\chi=32$ preserves 99.94\% of the image fidelity on average. 

We compute the encoding quantum circuits using the MPD and MPD+TNO algorithms. To significantly improve computational efficiency, we set $\chi_{\text{max}}=32$ (i.e., we restrict bond dimensions to grow to at most $\chi=32$). This implies that $F=0.9994$ is the maximum achievable encoding fidelity in this study. Additionally, we also implement the unitary synthesis method that maps an MPS of bond $\chi$ into a single sequential layer of at most $(\log_2(\chi)+1)$-qubit gates (rounding $\log_2{\chi}$ up to the nearest integer). The circuits computed by both methods are then decomposed into $CNOT$ and arbitrary single-qubit $U(3)$ gates using Qiskit, allowing a direct comparison of gate count and circuit depth.

\begin{figure}[H]
    \centering
    \begin{subfigure}[b]{0.45\textwidth}
        \centering
        \includegraphics[trim=0cm 0cm 0cm 0cm, clip, width=\textwidth]{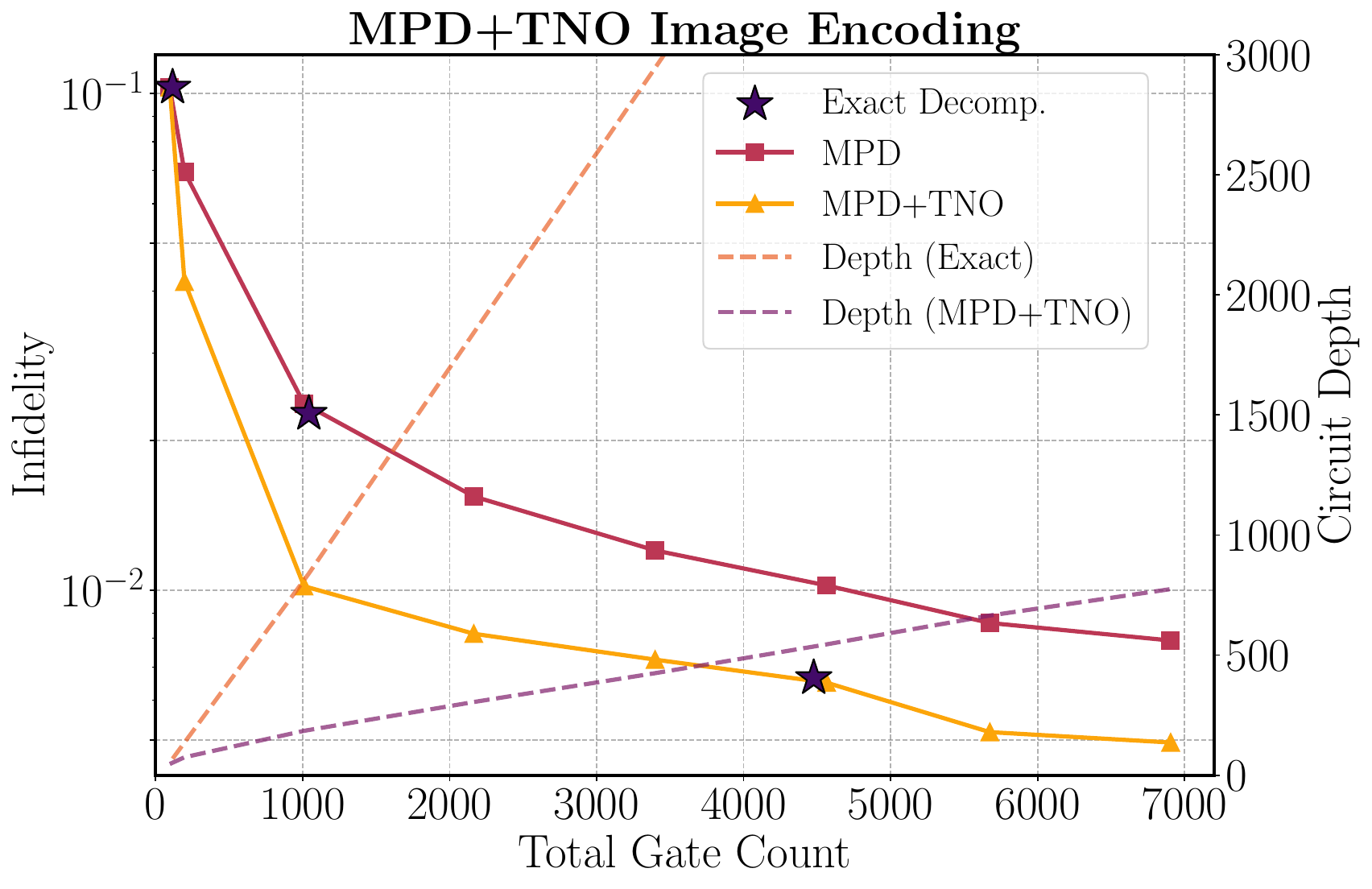}
        \centering
        \caption{}
        \label{fig:encoding_tno}
    \end{subfigure}%
    \hfill
    \begin{subfigure}[b]{0.5\textwidth}
        \centering
        \includegraphics[trim=0cm 0cm 0cm 0cm, clip, width=\textwidth]{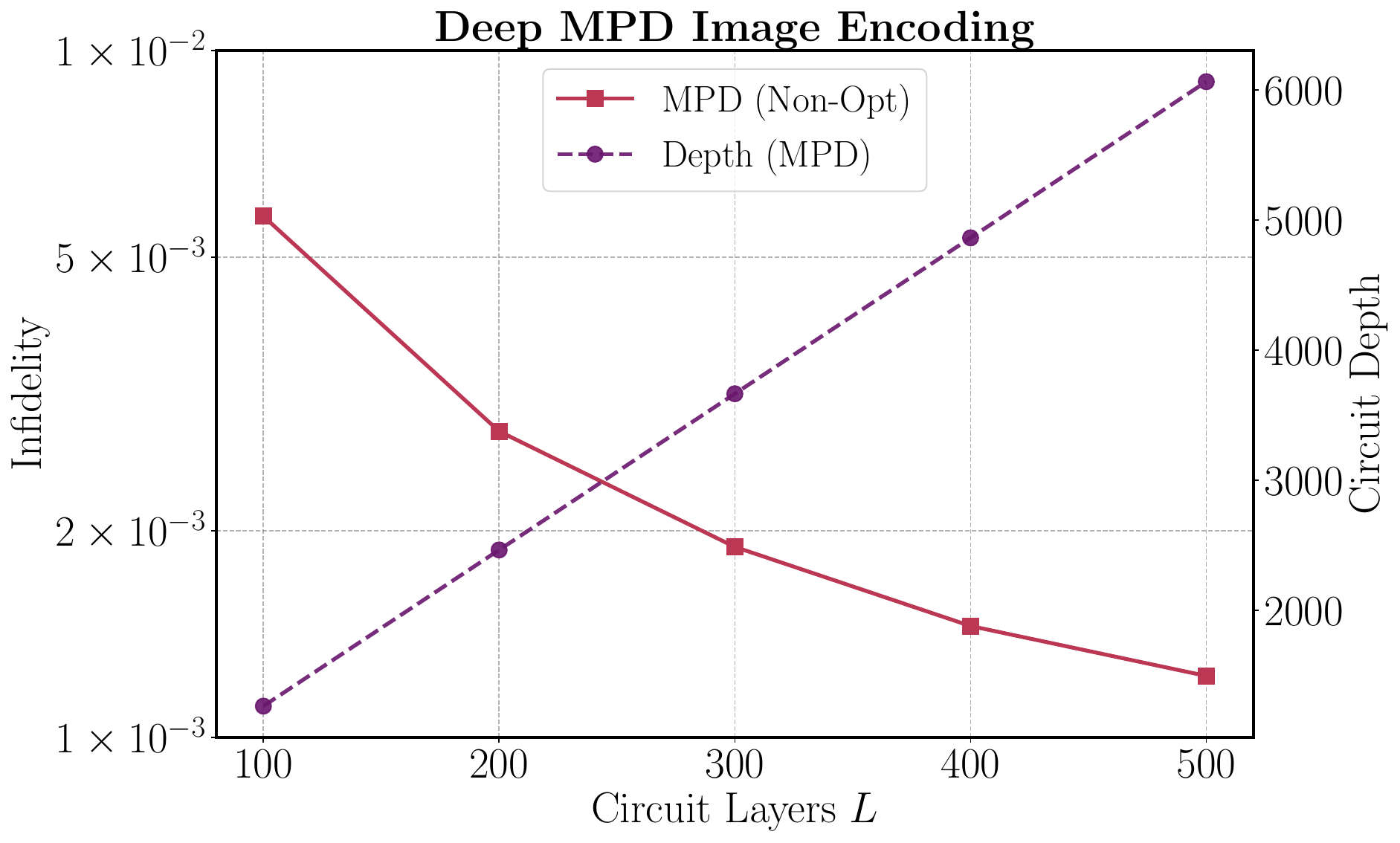}
        \centering
        \caption{}
        \label{fig:encoding_deep}
    \end{subfigure}
    \caption{Quantum encoding of a $128\times 128$ ChestMNIST image: (a) Comparing the MPD+TNO algorithm against the exact decomposition of a bond-$\chi$ MPS approximation with shallow-depth ($L=1$ to 60) circuits. The exact decomposition is the direct mapping of an optimal bond-$\chi$ MPS into a single sequential layer of at most $(\log_2{\chi}+1)$-qubit gates (rounding $\log_2{\chi}$ up to the nearest integer) before decomposition into 1- and 2-qubit gates is computed with Qiskit. This is computed for $\chi=2,4,8$. The MPD+TNO algorithm outperforms the exact decomposition in terms of total gate count for shallow circuits, and significantly reduces the total circuit depth. (b) The MPD algorithm without optimisation continues to converge even at deep circuit layers. The fidelity $F=0.9981$ achieved at $L=300$ layers approximately achieves the accuracy of an optimal $\chi=16$ encoding of the image ($F=0.9982$) with a circuit depth of 3665 gates. In comparison, the circuit depth achieved by an exact decomposition for $\chi=16$ is 15060 gates. }
    \label{fig:encoding_results}
\end{figure}

The encoding results comparing the MPD,MPD+TNO, and the exact decomposition approach are presented in Figure~\ref{fig:encoding_results}(a). The MPD+TNO circuits outperform the exact decomposition method for shallow layers (up to $\sim L=50$ or a total gate count of 4500), achieving higher fidelity encodings with lower total gate count. While the MPD algorithm initially matches the performance of the exact decomposition, it begins to converge more slowly, exhibiting marginally decreasing gains in fidelity with increasing layers. Surprisingly, as shown in Figure~\ref{fig:encoding_results}(b), the MPD algorithm does not plateau, but increases to improve fidelity even at deep layers ($L=100$ to 500). The output fidelities of the computed circuits are visualised in Figure~\ref{fig:subfig_encoded_images}. The output state vector corresponding to each circuit is computed by simulating the circuit in Qiskit, before being reshaped into the original image dimensions for visualisation.

There are two definitive advantages of the MPD+TNO algorithm over the exact decomposition in this problem. Most significantly, the MPD+TNO algorithm produces significantly \textit{shallower} circuit depths than the exact decomposition, even when the total gate count is similar. This is because the exact decomposition method requires all ($\log_2{\chi}+1$)-qubit gates to be applied in a strict \textit{sequential} order. This means that, after decomposition, the circuit depth is approximately the same as the total gate count. In comparison, the multi-layer sequential circuits of $\chi=2$ MPOs already exhibit significant parallelisation across all qubits, meaning that the circuit depth scales as $\mathcal{O}(\frac{1}{n}\times\text{Total Gate Count})$. Secondly, the exact decomposition produces significantly suboptimal circuits for bond dimensions $\chi$ that are not powers of 2. If $\chi$ is not a power of 2, then the bond-$\chi$ tensors must be padded into larger unitary gates, leading to deeper-than-necessary circuits after decomposition. At best, the circuit depth of the exact decomposition method could be halved by moving the orthogonality centre of the constructed MPS to the central tensor, which would allow both ends of the MPS to be generated in parallel.

In Figure~\ref{fig:encoding_results}(a), the exact decomposition is computed for $\chi=2,4,8$, which is achieved by the MPD+TNO with 1,3, and 40 layers, respectively. The MPD algorithm without optimisation achieves a fidelity of $F=0.9981$ at $L=300$ layers with a circuit depth of 3665 gates. This is approximately the accuracy of an optimal $\chi=16$ encoding of the image ($F=0.9982$). In comparison, the circuit depth achieved by an exact decomposition for $\chi=16$ is 15060 gates. However, with the number of variational circuit parameters scaling as $\mathcal{O}(nL)$, these deep circuits pose a computational burden for optimisation. This is supported by the inability of the MPD+TNO algorithm to outperform the exact decomposition for the $\chi=8$ encoding in terms of total gate count, likely indicating local minima in the optimisation process despite the MPD initialisation of parameters.

It is well-known that all quantum MPS of bond-$\chi$ with longer-range correlations can be exactly expressed in quantum circuits of $\mathcal{O}(n\chi^2)$ total gates \cite{malz_mps_encoding}. This is the circuit depth guaranteed by the exact decomposition approach, though with potentially suboptimal constant factors. However, the question of how many gates are required for a specific bond-$\chi$ MPS is not generally known. Here, we observe that the MPD+TNO algorithm performs similarly to the exact decomposition method for the image encoding problem. This is likely indicative that the total $\mathcal{O}(n\chi^2)$ gate count is necessary to express the entanglement entropy characteristic of the ChestMNIST image. This is in line with recent findings for the encoding of the Fourier coefficients of image data with sequential circuits, where a gate count scaling with a low-degree power of $\chi$ was observed \cite{Jobst2024efficientmps}.

The proposed MPS image encoding scheme performs surprisingly well, suggesting that the target images can be well-approximated by $\chi$-reduced MPS structures. We suggest that the MPS format exploits the locality of the entanglement structure of natural images, enabling it to form high-quality image approximations. This is a testament to the expressive power of the MPS. However, this encoding protocol represents a state near the boundary of what can be efficiently achieved by the MPD+TNO algorithm. Based on our findings, it is reasonable to assume that images with Schmidt coefficient spectra that do not obey the near-power law decay observed in Figure~\ref{fig:mps_compression_graphic}(b) would be outside of the expressible scope of the MPS structure for a bond dimension that can be efficiently encoded via the MPD+TNO algorithm. As argued in \cite{lu_tn_efficient_descriptions_classical_data}, more expressive tensor networks will be essential for capturing the structure of more complex images.

    

    

\begin{figure}[htb]
         \centering
        \includegraphics[trim=0cm 0cm 0cm 0cm, clip, width=0.8\textwidth]{New_Image_Encoding_Results/image_comparison_blue.png}
        \caption{An example target $128\times 128$ ChestMNIST alongside the reshaped image encoded by the MPD and MPD+TNO algorithms on $n=14$ qubits for various circuit layers. All MPD+TNO results were computed using 400 training iterations. Images were reconstructed by reshaping the simulated state vector into the shape of the original image.}
        \label{fig:subfig_encoded_images}
\end{figure}
    

\section*{6. Discussion and Conclusion}

While the $2^n$-dimensional Hilbert space represents an immense resource for computation, it imposes a dimensionality burden in the context of quantum state preparation. In turn, the absence of any general and efficient algorithm for arbitrary state preparation represents a severe bottleneck for quantum machine learning, linear algebra, and many other significant algorithms. Resultingly, we confined our scope to the design of general algorithms for the state preparation of structured states. Motivated by the avoidance of resource-intensive circuits for quantum arithmetic, lack of ancillary qubits, and in-built dimensionality reduction, we adopted an MPS-based framework for the approximate state preparation of structured data centred around the preparation of discretised functions and images. We utilised the entirely classical MPD and MPD+TNO algorithms as the focal points of this investigation.

Every quantum state produced by $L$ sequential circuit layers of $\chi=2$ MPOs can be expressed as an MPS with bond dimension at most $\chi_{\text{max}}=2^L$. Naively, one might assume that this implies that all bond dimension $\chi$ MPS can be expressed in circuits of depth $O(n\log_2{L})$. However, only a small subset of bond $\chi$ MPS exist in this subspace. Since a bond-$\chi$ MPS carries $\mathcal{O}(n\chi^2)$ independent real parameters, a circuit of $\mathcal{O}(n\chi^2)$ quantum gates is necessary to express all bond-$\chi$ MPS. In general, the number of sequential circuit layers necessary to represent a specific MPS as a function of its bond dimension is not known \cite{malz_mps_encoding}. Hence, we are motivated by the potential for the MPD algorithm to outperform the exact MPS decomposition method of \cite{schon_sequential_MPS_generation}, which is guaranteed to output a circuit of depth $\mathcal{O}(n\chi^2)$ (with potentially significant factors hidden by big-O notation) via the targeted MPD+TNO preparation strategy.

In the context of the state preparation of functions, the MPD algorithm introduced in \cite{ran} was previously shown to prepare smooth functions \cite{holmes_smooth_diff_functions} and Gaussian distributions in the probabilities \cite{iaconis}. We extended these results to a broader class of functions, including those with irregular behaviour, such as simple discontinuities in piecewise polynomials and steep gradients exemplified by the root and logarithmic functions. This extension was primarily supported by the addition of TNO to the MPD algorithm, which improved the fidelity of the prepared state for any given number of circuit layers. This improvement was most pronounced in cases where the MPD algorithm performed poorly, i.e. when the iterative $\chi=2$ disentangling layers only minimally decreased the entanglement in the state. 

Many of the prepared simple functions have immediate applications. For example, we can exactly prepare the `hockey-stick' function $f(x)=\max{(0,x-K)}$ for some constant $K$, which is used as the boundary condition when solving the Black-Scholes PDE \cite{yusen_portfolio_optimisation}. Since this state can be prepared exactly with a single layer of $U(4)$ gates, this is likely the most resource-efficient approach to preparing similarly straightforward functions. For linear algebra problems, where very high accuracy is essential, we identify that the MPD and MPD+TNO approaches can encode up to low-degree piecewise polynomials with approximately 10 intervals with fidelity exceeding 99.99\% using shallow, linear-depth circuits. More complex structures can be approximately encoded using similarly shallow-depth circuits if a more significant error in the encoded state is permissible.

We also applied the MPD and MPD+TNO to the problem of quantum image encoding. The proposal of MPS-based image encoding has been previously considered in \cite{Jobst2024efficientmps}, where it was demonstrated that the MPS format could efficiently prepare the discretised Fourier representation of images if the coefficients exhibit an exponential or power-law decay rate. It was also considered in \cite{compressed_prep_for_QML}, in which the exact state preparation circuit is compressed analytically via MPS, before the MPS is mapped directly into a quantum circuit via unitary synthesis. In this work, we use the efficient classical methods to approximately map $\chi$-reduced MPS representations of ChestMNIST pixel values into a quantum circuit, demonstrating a novel approach to quantum image encoding with shallow-depth circuits. We showcase that the direct encoding via MPD+TNO produces significantly shallower-depth circuits than the direct decomposition, owing to an increased parallelisation. Additionally, the direct encoding of pixel values avoids the requirement of inverse transforms in the case of the initial preparation of discrete Fourier or discrete wavelet coefficients. 

Intuitively, however, as the complexity in the target state grows and it starts to approximate unstructured noise, even small reductions in the bond dimension will significantly detract from the fidelity of the approximated image. In the case of straightforward discretised functions, the appreciable structure of the target state enables highly efficient MPS approximations, which we encode with accuracy exceeding 99.99\%. On the contrary, natural images can only be roughly approximated with the $\chi$-reduced MPS format, owing to the locality of structure within them. More complicated images with nuanced, long-range structures, however, necessitate the adoption of more expressive tensor networks and deeper quantum state preparation circuits. Future work may consider the application of tensor structures more adept at capturing hierarchical, multi-scale entanglement structures. For example, recent work has considered applying more general tree tensor networks for the state preparation of multivariate normal distributions \cite{manabe2024statepreparationmultivariatenormal}. A tree-based or 1D MERA approach may be a potential pathway to improving the performance of the proposed tensor network quantum image encoding scheme.

Interestingly, the authors in \cite{compressed_prep_for_QML} identified the pseudo-randomness of the MPS decomposition as a potential form of regularisation during training. This finding is complemented by evidence for this implicit regularisation effect in a range of entirely classical tensor network machine learning models \cite{NIPS2016_5314b967_ctnml,Singh_Bhatia_2024_ctnml,10.1162/neco_a_01202_ctnml}. Furthermore, there is ample opportunity for future research in the design of task-aware MPS encodings. For example, MPS have been used to pre-train quantum machine learning models \cite{Dborin_2022,Rudolph2023}, in which the target state of the MPS-based preparation algorithm represents the MPS computed as the output of a low-rank classical tensor network optimisation problem. This classical tensor network optimisation solution can theoretically be ``quantised" via the addition of new quantum gates to increase the expressivity of the quantum circuit, while reducing the effects of barren plateaus through the pre-optimised classical MPS parameters. 

An alternative and promising approach to quantum state preparation not considered by this article is based on quantum signal processing (QSP), especially the Quantum Eigenvalue Transform (QET) \cite{mcardle2022quantumstatepreparationcoherent,gonzalez}. The QSVT approach works to implement degree-$d$ polynomials on $n$ qubits with circuit depth $\mathcal{O}(nd)$, enabling the preparation of polynomials with degree $d=\mathcal{O}(\text{poly}(n)$. In \cite{gonzalez}, the authors use a non-linear transformation of amplitudes method introduced in \cite{rattew2023nonlineartransformationsquantumamplitudes}, which enables polynomial transformations to the amplitudes of any efficiently preparable quantum state. This approach requires $\mathcal{O}(1)$ ancilla qubits and $\mathcal{O}(1)$ calls to a controlled-$U_S$ circuit, where $U_S$ is the initial state preparation circuit. We note that the MPS-based preparation scheme permits much shallower-depth circuits consisting of only local gates without any ancilla qubits, which is more resource-efficient than the QET approach. However, future work may more directly compare the resource costs of the exact MPS encoding scheme in \cite{schon_sequential_MPS_generation} with the QET approach for high-degree polynomials.

The authors in the QET approach in \cite{gonzalez} introduce a novel state preparation circuit for preparing the linear function $f(x)$ using the Discrete Hadamard-Walsh Transform (DHWT), prepared with a $\mathcal{O}(n)$-depth circuit consisting of controlled-$R_Y$ gates. We note that the MPS-based approach can be used to prepare the linear function state exactly with just a single sequential layer, owing to the existence of an exact $\chi=2$ MPS representation of the state. Hence, the MPS preparation of $f(x)=x$ could viably be used in place of the DHWT encoding approach, simplifying the circuit structure. 

In conclusion, the MPD and MPD+TNO algorithms are exceptionally well-suited to near-term NISQ devices with applications ranging from the state preparation of boundary conditions for PDEs, probability distributions for quantum Monte Carlo, and approximate images for quantum image processing and classification. Alongside the extension to QET-based state preparation, benchmarking of the explored algorithms in comparison to other recently developed approximate MPS preparation schemes based on the renormalisation-group transformation \cite{malz_mps_encoding} and adaptive circuits \cite{constant_depth_mps_smith} is a key area of future research. Notably, MPS-based state preparation techniques can be entirely efficiently simulated using classical devices, meaning that ``quantum advantage" cannot be derived directly from assumptions of state preparation. The pressing question remains: \textit{for what applications does this permit a quantum advantage?} In any case, the explored algorithms are highly versatile and well-suited to the current stage of quantum development, enabling implementation on existing quantum devices for further experimentation and testing.

\section{Acknowledgements}
The authors wish to thank Matthaus Zering for valuable and insightful discussions, and to acknowledge continued support from the Pawsey Supercomputing Research Centre.

\bibliography{references}

\end{document}